%% file: manuscript.tex
\IfFileExists{/dev/null}{\immediate\write18{sh makeLinks.sh}}{\immediate\write18{makeLinks.bat}}

\documentclass[a4paper,journal,twocolumn,10pt,final]{IEEEtran}

\makeatletter
\def\subsubsection{%
	\@startsection
	{subsubsection}                 
	{3}                             
	{\z@}                           
	{2.5ex plus 1.5ex minus 1.5ex}  
	{1ex plus .5ex minus 0ex}     
	{\normalfont\normalsize\itshape}
}
\makeatother

\usepackage[none]{hyphenat}
\usepackage[utf8]{inputenc}
\usepackage[T1]{fontenc}
\usepackage{textcomp}
\usepackage{scalefnt}
\usepackage{amsmath,amsfonts,amssymb,amsbsy,amstext}

\usepackage{mathtools}          
\usepackage{cmap}               
\usepackage[dvipsnames,svgnames,x11names]{xcolor}   
\usepackage{varwidth}

\let\originalleft\left
\let\originalright\right
\renewcommand{\left}{\mathopen{}\mathclose\bgroup\originalleft}
\renewcommand{\right}{\aftergroup\egroup\originalright}

\usepackage{graphicx}
\graphicspath{{.}{.\figs}}
\usepackage{placeins} 

\usepackage{multirow}
\usepackage{tabularx}
\usepackage{booktabs}
\newcolumntype{C}{>{\centering\arraybackslash}X}
\newcolumntype{R}{>{\flushright\arraybackslash}X}
\newcolumntype{L}{>{\flushleft\arraybackslash}X}
\newcolumntype{P}{>{\centering\arraybackslash} p{0.5\linewidth}}

\usepackage[%
style=ieee,
backend=biber, 
]{biblatex}

\defbibheading{bibliography}[\refname]{\section*{#1}}

\makeatletter
\g@addto@macro{\UrlBreaks}{\UrlOrds}
\makeatother

\usepackage[version=2]{acro}

\NewDocumentCommand{\acro}{m o m o}
{%
	\IfValueTF{#2}{%
		\IfValueTF{#4}{%
			\DeclareAcronym{#1}{short={#2},long={#3},#4}
		}{%
			\DeclareAcronym{#1}{short={#2},long={#3}}
		}
	}{%
		\IfValueTF{#4}{%
			\DeclareAcronym{#1}{short={#1},long={#3},#4}
		}{%
			\DeclareAcronym{#1}{short={#1},long={#3}}
		}
	}
}

\input{acronymsLowerCase.tex}

\usepackage{siunitx}
\usepackage{datetime}
\usepackage{url}

\usepackage[caption=false,font=footnotesize]{subfig}

\usepackage{algorithm}
\usepackage{algorithmicx}
\usepackage{algpseudocode}

\usepackage{epstopdf}

\usepackage[hidelinks=true]{hyperref}

\usepackage[final]{changes} 
\definechangesauthor[name={Victor F. Monteiro}]{vfm}
\definechangesauthor[name={Tarcisio F. Maciel}]{tfm}
\setaddedmarkup{{\color{blue}#1}}
\setdeletedmarkup{{\color{red}\sout{#1}}}

\usepackage[referable,flushleft]{threeparttablex}
\AtBeginEnvironment{tablenotes}{\footnotesize} 

\input{math.tex}

\def\tx{\mathrm{tx}}

\newcommand{\SecRef}[2][]{Section#1~\ref{#2}}
\newcommand{\FigRef}[2][]{Fig.#1~\ref{#2}}

\newcommand{\TabRef}[2][]{Table#1~\ref{#2}}

\usepackage{standalone}
\usepackage{shapepar}

\newcommand{\includetikzplot}[1]{%
	\includegraphics{figs/#1.pdf}
}

\AtBeginDocument{%
	\abovedisplayskip 1.5ex plus 4pt minus 2pt
	\belowdisplayskip \abovedisplayskip
	\abovedisplayshortskip 0pt plus 4pt
	\belowdisplayshortskip 1.5ex plus 4pt minus 2pt
}

\begin{document}
\title{Beam Squinting Compensation: \\An NCR-Assisted Scenario}

\author{Diego A. Sousa, Fco. Rafael M. Lima, \IEEEmembership{Senior Member, IEEE}, Victor F. Monteiro, \IEEEmembership{Member, IEEE}, \\ Tarcisio F. Maciel, \IEEEmembership{Senior Member, IEEE}, and Behrooz Makki, \IEEEmembership{Senior Member, IEEE}
\thanks{Behrooz Makki is with Ericsson Research, Sweden. The other authors are with the Wireless Telecommunications Research Group (GTEL), Federal University of Cear\'{a} (UFC), Fortaleza, Cear\'{a}, Brazil. Diego A. Sousa is also with Federal Institute of Education, Science, and Technology of Cear\'{a} (IFCE), Paracuru, Brazil. This work was supported by Ericsson Research, Sweden, and Ericsson Innovation Center, Brazil, under UFC.51 Technical Cooperation Contract Ericsson/UFC. The work of Victor F. Monteiro was supported by CNPq under Grant 308267/2022-2. The work of Tarcisio F. Maciel was supported by CNPq under Grant 312471/2021-1. The work of Francisco R. M. Lima was supported by FUNCAP (edital BPI) under Grant BP5-0197-00194.01.00/22.}
}

\maketitle

\begin{abstract}
\Ac{mmWave} and sub-THz communications, foreseen for \ac{6G}, suffer from high propagation losses which affect the network coverage. %
To address this point, smart \deleted{electromagnetic }entities such as \acp{NCR}\deleted{ and \acp{RIS}} have been considered as cost-efficient solutions \added{for coverage extension}. %
\Acp{NCR}, which have been \replaced{standardized in \acl{3GPP} Release 18}{proposed in Release~18 from \acl{3GPP}}, are \acl{RF} repeaters with beamforming capability controlled by the network through side control information. %
Another challenge raised by the adoption of high frequency bands is the use of large bandwidths. %
\replaced{Here}{To address this point}, a common configuration is to divide a large frequency band into multiple smaller subbands. %
In this context, \replaced{we}{the present work} consider\deleted{s} a scenario with \acp{NCR} where signaling related to measurements used for \acl{RRM}\deleted{, e.g., link adaptation,} is transmitted in one subband centered at frequency $f_c$ and data transmission is performed at a different frequency $f_c + \Delta f$ based on the measurements taken at $f_c$. %
\replaced{Here, a challenge}{A challenge that needs to be overcome to enable this subband signaling operation} is that the array radiation pattern can be frequency dependent and, therefore, lead to beam misalignment, called beam squinting. %
\replaced{We}{Considering this, we} characterize beam squinting in the context of subband operation and propose a solution where the beam patterns to be employed at a given subband can be adjusted/compensated to mitigate beam squinting. %
\replaced{Our results show}{Based on simulation results, we conclude} that\added{,} without \deleted{a }compensation, the perceived \ac{SINR} and so the \deleted{\ac{UE} }throughput can be substantially decreased\added{ due to beam squinting}. %
\deleted{For instance, the \ac{SINR} of the 10-percentile decreases by 20 dB compared to the case when signaling and data are transmitted at the same frequency.} %
\replaced{However}{On the other hand}, with our proposed compensation \added{method}, the system is able to support \ac{NCR} subband signaling operation with similar performance as if signaling and data were transmitted at the same frequency. %
\end{abstract}

\begin{IEEEkeywords}
	Beam squinting, subband, \acs{NCR}\added{, \acs{RIS}}.
\end{IEEEkeywords}

%
\IEEEpeerreviewmaketitle
\acresetall

\section{Introduction}
\label{CHP:Beam_Squint_SEC:Intro}

One of the main differences of the \ac{5G} of wireless cellular telecommunications systems compared to previous generations is the use of \ac{mmWave} spectrum. %
The use of even higher frequency bands is also planned for \ac{6G} where studies and experiments for communications in sub-THz spectrum are taking place~\cite{Shehata2023, Rajatheva2021}. %
This has at least two important impacts: i) the need for a dense deployment of different types of access points such as \ac{IAB}, \acp{NCR} or \acp{RIS}, due to the high propagation loss in these bands compared to sub-\SI{6}{GHz} bands; and ii) the usage of large bandwidths~\cite{Rajatheva2021}. %

Concerning the increased number of \replaced{base stations}{BSs} per square meter in \ac{5G} and \replaced{beyond}{next generation of mobile networks}, it demands improved backhaul infrastructure. %
Fiber is the default choice, if available, to provide backhaul connection from the \acp{gNB} to the \ac{CN}. %
Nevertheless, deploying fiber for backhaul connection may be costly and take time, e.g., due to trenching and installation, and can even be impossible in, e.g., historical places where trenching is not an option. %
To address the need of coverage enhancement in \ac{5G} and decrease the costs related to the deployment of conventional \acp{gNB}, new network nodes have been considered, e.g., \acp{NCR}~\cite{Guo2022,Silva2023}, \ac{IAB}~\cite{Monteiro2022,Madapatha2020} and \acp{RIS}~\cite{Chen2023}. %

\begin{figure}[!t]
	\centering
	\includegraphics[width=0.95\columnwidth]{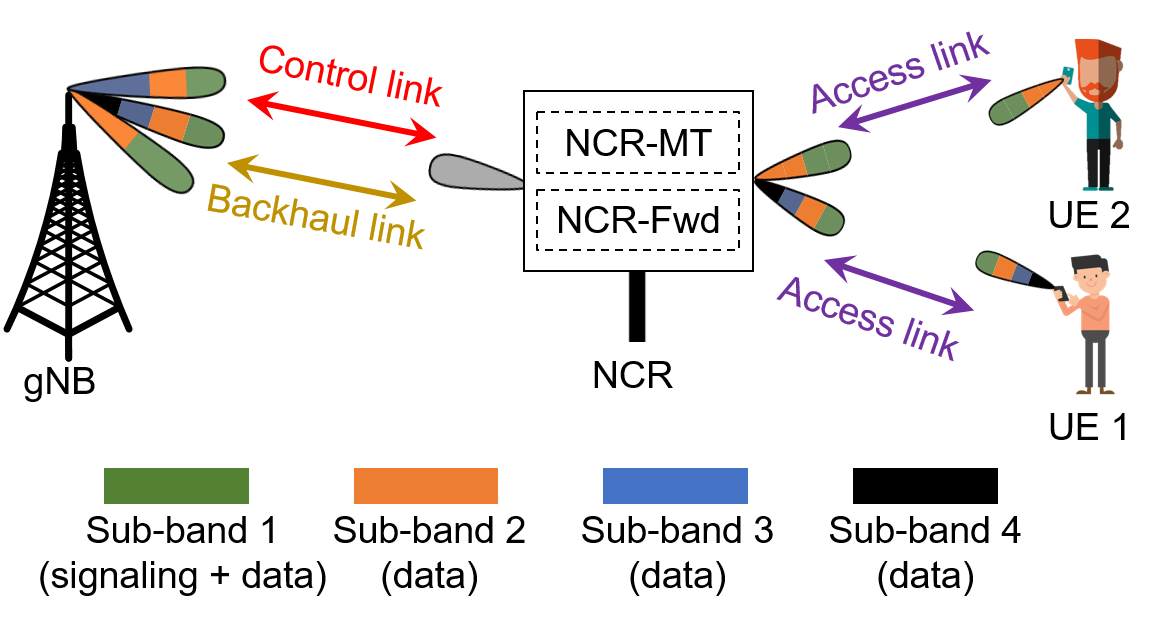}
	\caption{NCR supporting sub\deleted{-}band operation.}
	\label{CHP:Beam_Squint_FIG:proposal}
\end{figure}

\acp{NCR} have been \replaced{standardized}{introduced} by the \ac{3GPP} in \ac{5G}/\ac{NR} standards Release~18~\cite{3gpp.38.867}. %
While \ac{RF} repeaters, i.e., traditional repeaters, simply \ac{AF} received signals, \acp{NCR} can also receive side control information from a \ac{gNB} and perform beamforming~\cite{Lin2022}. %
As illustrated in~\FigRef{CHP:Beam_Squint_FIG:proposal}, an \ac{NCR} can be split into two parts: \ac{NCR}-\ac{MT} and \ac{NCR}-\ac{Fwd}. %
The first one is responsible for controlling the \ac{NCR} operation via a control link, based on the Uu interface, that exchanges  the side control information between \ac{gNB} and \ac{NCR}. %
The \ac{NCR}-\ac{Fwd} is responsible for the \ac{AF} relaying in access and backhaul links, \replaced{shown in Fig. \ref{CHP:Beam_Squint_FIG:proposal}, and is}{being} controlled via the side control information received by the \ac{NCR}-\ac{MT}. %

Concerning the use of a large bandwidth in \ac{5G}/\ac{NR}, in a standard network configuration, a \ac{UE} is required to perform more measurements to estimate the quality of a link in the whole band. %
As a consequence of the increase in the amount of required \ac{UE} measurements, \ac{UE} battery consumption is also expected to increase. 
Another consequence of having a large bandwidth is that \acp{UE} may have different capabilities and support different bandwidths. %
To address this aspect, a common configuration in \ac{mmWave} is to divide a large frequency band into multiple smaller subbands~\cite{3GPP-R1-2207680}. %

In this context, we consider the topics of \ac{NCR} and subband operation in order to deal with: i) the need of coverage enhancement; and ii) the challenge of using large bandwidths. %
\replaced{Figure}{More specifically,} \ref{CHP:Beam_Squint_FIG:proposal} illustrates the considered setup. 
In this example, the system bandwidth is split into four subbands. %
\ac{UE}~$1$ supports all subbands, while \ac{UE}~$2$ only supports subbands $1$ and $2$. %
Instead of transmitting signaling information, i.e., \ac{CP}, in the whole band, as in the state-of-the-art, we consider  the possibility of transmitting signaling related to measurements used for \ac{RRM}, e.g., link adaptation, in just one subband, e.g., subband~$1$ in \FigRef{CHP:Beam_Squint_FIG:proposal} (which is supported by both \acp{UE}), while data transmission, i.e., \ac{DP}, may be performed in \replaced{different}{the other} subbands based on the measurements performed at subband~$1$. %

A challenge that needs to be overcome to enable subband signaling operation with repeater nodes, e.g., \acp{NCR} and \acp{RIS}, is that the array radiation pattern can be frequency dependent and, therefore, beam misalignment can occur due to the effect of \emph{beam squinting}. %
Thus, reusing the transmitter parameters, e.g., beamforming filters, defined for a given subband in other subbands may lead to performance degradation. %
Beam squinting has been the main subject of some articles in the literature in different contexts. %
In~\cite{Cai2016}, the authors studied the impact of beam squinting in \ac{ULA} with analog beamforming. %
In order to reduce the impact of beam squint in a wideband scenario, the authors suggested a codebook design to enforce a minimum array gain in the whole bandwidth. %

Beam squint for wideband hybrid beamforming was considered in~\cite{Gao2021,Tan2019,Dai2022,Wan2021,Zhai2020_1}. %
In \cite{Wan2021}, analog and digital precoders/combiners were designed to maximize spectral efficiency. %
In \cite{Gao2021}, the authors proposed two beamforming techniques to combat beam squinting where analog and digital precoders/combiners were designed. %
Firstly, the authors proposed a virtual sub-array architecture that performs a beam broadening that provides a more uniform gain over the operating frequency range. %
The other solution is based on the addition of a few true-time-delay lines to compensate for differences in time delays among different antennas. %
True-time-delay lines were also considered in \cite{Tan2019,Dai2022,Zhai2020_1} associated with analog phase shifts in order to make phase shifts frequency dependent. %
Although the use of true-time-delay lines improves performance by mitigating beam squinting, they increase energy consumption and hardware cost compared to conventional designs. %

Beam squinting was also studied for \ac{RIS}-assisted networks~\cite{Hao2023,Ma2021}. %
In~\cite{Hao2023}, the authors analyzed the impact of beam squinting in near and far fields. %
To mitigate beam squinting, the authors proposed a delay adjustable metasurface architecture that is equivalent to the use of true-time-delay lines in phased antenna arrays. %
Therefore, signals reflected by different \ac{RIS} elements are submitted to adjustable delays in order to deal with the beam gain loss caused by beam squinting. %
In~\cite{Ma2021}\added{,} the authors focused on channel estimation methods for wideband \ac{RIS}\replaced{, and}{. Here, the authors} proposed a method for the frequency domain channel response estimation based on cross-entropy theory. %

In this paper, we study the problem of beam squinting in \ac{NCR}-assisted networks. %
Although our proposed method can be applied to general \ac{MIMO} scenarios, \acp{NCR} can be deployed in crowded areas where the distances to the terminals are not high. %
Thus, \acp{UE} are not necessarily aligned to the antenna boresight of \ac{NCR} panels, which increases the beam squinting effect. %
Motivated by this, we characterize the beam squinting phenomenon in the context of subband operation and propose a beam squinting compensation \added{method} to enable the aforementioned subband operation illustrated in \FigRef{CHP:Beam_Squint_FIG:proposal}. %
Then, we present a system-level performance evaluation where we show the impact of beam squint in the aforementioned scenario. %
Beam squinting leads to both main lobe misalignment in the radiation pattern as well as higher interference in the system\added{,} which decreases \ac{SINR}. %
Furthermore, we show that our proposed compensation method is capable of keeping a similar system performance in terms of throughput and \ac{SINR} for different carrier frequency displacements by avoiding the undesired effects of beam squinting. %

Our work is different from those in the literature because previous works \cite{Cai2016,Gao2021,Tan2019,Dai2022,Wan2021,Zhai2020_1,Tan2019,Dai2022,Zhai2020_1,Hao2023,Ma2021} in general had the objective of increasing the operation bandwidth of the antenna arrays by mitigating beam squinting. %
On the other hand, our objective is to reduce the beam squinting in order to transmit in a second subband based on the collected information, e.g., \ac{CSI}, and transmission parameters, e.g., precoders, collected/defined from/to a first subband. %
Moreover, previous works in general focused on simple evaluation scenarios involving few network nodes and, thus, are not capable of showing the achievable gains of their proposals from a system-level perspective. %
\added{Finally, our proposed beam squinting compensation method has not been presented before.} %

The present work is organized as follows. %
First, in \SecRef{CHP:Beam_Squint_SEC:Bea_Squ_Hig_Lev_Cha}, we provide a high-level discussion on the problem of beam squinting. %
Then, in \SecRef{CHP:Beam_Squint_SEC:Sys_Mod}, we present a generic model for antenna array and discuss the array beam pattern for different frequencies. %
In \SecRef{CHP:Beam_Squint_SEC:Prob_Def_Prop_Sol}, we analytically show the phase deviation problem and present our proposed solution. %
\SecRef{CHP:Beam_Squint_SEC:Perf_Eval} presents a performance evaluation of our proposed method based on simulations. %
Finally, \SecRef{CHP:Beam_Squint_SEC:Conclusion} presents the conclusions of this work. 

\section{Beam squinting: high-level characterization}
\label{CHP:Beam_Squint_SEC:Bea_Squ_Hig_Lev_Cha}

Together with the antenna element radiation pattern, the antenna array geometry is one of the defining characteristics of the performance of a system employing \ac{MIMO} antennas. When designing a \ac{MIMO} system, regular/uniform geometries are usually applied (despite not mandatory) so that most \ac{5G} communication system assume the usage of
standard linear arrays or standard rectangular arrays. By \textit{standard} herein, one means that the antenna element spacing $d$ in the antenna arrays is normally set to $\frac{\lambda_c}{2}$, where $\lambda_c$ is the wavelength of the central carrier frequency $f_c$ of the system being designed. %
With such spacing, signals impinging as a planar wavefront on the antenna arrays are spatially sampled in an optimum way according to Nyquist's theorem. %
As a consequence, several important properties, such as beam orthogonality, can be exploited by the \ac{MIMO} systems to spatially filter signals\replaced{,}{as} to reinforce components of interest and/or suppress interfering ones. %

These assumptions are valid for narrow band systems in which the system bandwidth $B$ around $f_c$ is very small compared to the carrier frequency itself, i.e.\added{,} $f_c \gg B$. In this case, frequencies $f = f_c \pm \frac{B}{2}$ lead to wavelengths $\lambda \approx \lambda_c$. However, with the advent of \ac{mmWave} and possible use of sub-THz in \ac{6G}, broadband systems are expected to operate with much larger bandwidths than their predecessors and its contemporaneous sub-\SI{6}{GHz} version. %
In this sense, systems designed for a center carrier frequency $f_c = f_1$, i.e., with antenna arrays designed considering $\lambda_c = \lambda_1$, but operating with multiple subbands, such as \ac{5G} \ac{NR}, might suffer from spatial aliasing. %
This may occur because, for a subband at frequency $f_2 = f_1 + \Delta f, \Delta f > 0$, and having a wavelength $\lambda_2 < \lambda_1$, the antenna element spacing will be perceived at the frequency $f_2$ as an antenna array designed with a suboptimal antenna element spacing $\frac{\lambda_1}{2}$ larger than the ideal one $\frac{\lambda_2}{2}$ for $f_2$. %
In the case of $\Delta f < 0$, the same effect would take place but the antenna element spacing would be smaller than the ideal one. %

If the antenna element spacing is larger than the ideal one (i.e., $\frac{\lambda_i}{2} > \frac{\lambda_c}{2}$ for frequency $f_i$), the expected spatial aliasing effect is a ``compression'' in the beam space generating narrowed grating lobes. %
On the other hand, if the antenna element spacing is smaller ($\frac{\lambda_i}{2} < \frac{\lambda_c}{2}$) than the ideal one, grating lobes still appear, however as ``widened'' beams~\cite[Sec.~2.4.1.2]{Trees2002}. %
In both cases, for a certain angular position associated with a beam at the center carrier frequency $f_1$, there will be angular shifts for the actual beam observed at frequency $f_2$. %
This change in how spatial signals and beams are viewed by the antenna array depending on the frequency (or subband) is also termed beam squinting. %
Notice that, often the beam associated with a \ac{UE} might represent the only angular information about that \ac{UE}. %
Therefore, angular shifts of beams may strongly affect the link quality of the \acp{UE} associated with those beams. %

\begin{figure}
	\subfloat[Pointing direction close to the antenna boresight.]{%
		\centering
		\includetikzplot{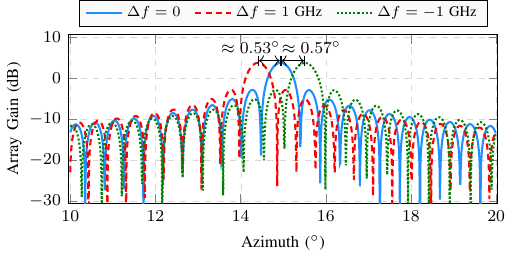}
		\label{FIG:Beam_Squint_SEC:beam_squint_repr:boresight}
	}
	
	\subfloat[Intermediate pointing direction.]{%
		\centering
		\includetikzplot{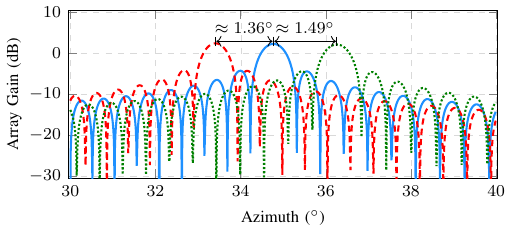}
		\label{FIG:Beam_Squint_SEC:beam_squint_repr:intermediate}
	}
	
	\subfloat[Pointing direction close to the antenna endfire.]{%
		\centering
		\includetikzplot{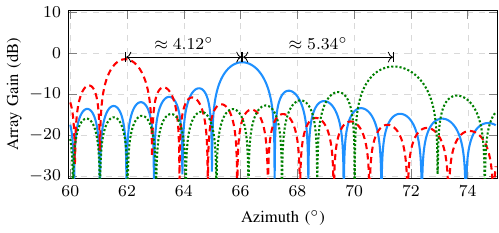}
		\label{FIG:Beam_Squint_SEC:beam_squint_repr:endfire}
	}
	\caption{Beam squinting representation considering a \ac{ULA} with 256 elements placed on the $y$-axis at $f_c = 28$ GHz
		and a frequency offset $\Delta f = \pm 1$ GHz and beams pointing toward different directions.}
	\label{FIG:Beam_Squint_SEC:beam_squint_repr}
\end{figure}

In \FigRef{FIG:Beam_Squint_SEC:beam_squint_repr}, we illustrate the beam squinting phenomenon for a \added{transmitter equipped with a} 256-linear antenna array with elements spaced by half wavelength assuming a center frequency $f_c=28$ GHz. %
In this figure, we show the antenna array radiation pattern for a beam designed (with a proper set of phase shifts) for the operating frequency $f_c$ (solid blue curve), and the array radiation patterns for the same phase shift configuration but at frequencies $f_c + \Delta f$ and $f_c - \Delta f$ (dashed-red and dotted-green curves, respectively) assuming $\Delta f = 1 \text{ GHz}$. %
Figure \ref{FIG:Beam_Squint_SEC:beam_squint_repr} is composed of three subfigures, i.e., \ref{FIG:Beam_Squint_SEC:beam_squint_repr:boresight}, \ref{FIG:Beam_Squint_SEC:beam_squint_repr:intermediate} and \ref{FIG:Beam_Squint_SEC:beam_squint_repr:endfire}, where a beam was designed at frequency $f_c$ for an impinging wavefront direction close to the antenna boresight, another intermediate direction, and an impinging wavefront direction close to the antenna array endfire, respectively. %

From \FigRef{FIG:Beam_Squint_SEC:beam_squint_repr}, it is clear that the beam squinting, i.e., the beam deviation, increases with $\Delta f$ and the angle which the beam was designed to point to relative to the antenna array boresight. %
For example, while the beam deviation for a beam pointing closely to the array boresight at $f_c+\Delta f$ was $0.57^{\text{o}}$, the beam deviation changes to $5.34^{\text{o}}$ for a beam at the same frequency pointing to an angle close to the array endfire. %
In the context of \acp{NCR}, this can be an issue since \acp{NCR} are usually deployed close to the \acp{UE} and so they are not necessarily aligned to the array boresight direction, being served by beams close to the array endfire. %

Another aspect that can be confirmed in \FigRef{FIG:Beam_Squint_SEC:beam_squint_repr} is the beam widening and compression when the operating frequency is lower than or greater than $f_c$, respectively. %
For example, in \FigRef{FIG:Beam_Squint_SEC:beam_squint_repr:endfire}, the \ac{HPBW} for the beams at frequencies 28~GHz, (28 + 1) GHz and (28 - 1) GHz are $1.33^\text{o}$, $1.11^\text{o}$ and $1.99^\text{o}$, respectively. %

\begin{figure}
	\centering
	\includetikzplot{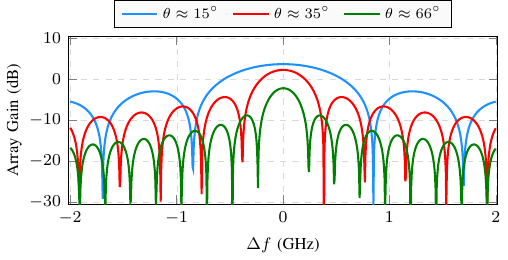}
	\caption{Beam squinting representation considering a \ac{ULA} with 256 elements placed on the $y$-axis at $f_c = 28$ GHz.}
	\label{FIG:Beam_Squint_SEC:beam_squint_repr_by_offset}
\end{figure}

In \FigRef{FIG:Beam_Squint_SEC:beam_squint_repr_by_offset}, we present another perspective of the beam squinting problem assuming the same antenna configuration as in \FigRef{FIG:Beam_Squint_SEC:beam_squint_repr}. %
However, the figure presents the array gain in dB versus the frequency deviation, $\Delta f$, assuming different beams whose main lobe points to azimuths $15^{\text{o}}$ (blue curve), $35^{\text{o}}$ (red curve) and $66^{\text{o}}$ (green curve) which represent examples of beams pointing to boresight, an intermediate direction and endfire of the antenna array, respectively. %
From this figure we observe in a more direct way the impact of the frequency deviation on the array gain loss due to the beam squinting phenomenon. %
Furthermore, as the absolute value of frequency deviation increases, the array gain decreases differently depending on the main lobe direction of the beams. %
The array gain of beams close to the boresight (blue curve, for example) requires a higher frequency deviation in order to present a significant loss (due to beam squinting). %
However, beams near the endfire of the antenna array (green curve, for example) are more sensitive to the frequency deviation. %

Beam squinting was also discussed analytically in~\cite{Garakoui2011}, where the authors derived an equation that shows the beam squinting for a linear array with equally spaced antenna elements assuming ideal phase shifters. %
The beam squinting was defined as the change in the beam main lobe direction, $\Delta\theta$, in terms of the original beam main lobe direction, $\theta_1$, the original operation frequency, $f_1$, and the frequency deviation, $\Delta f$. %
This equation is given by %

\begin{equation}\label{eq_delta_f}
	\Delta\theta = - \tan\left( \theta_1 \right)\frac{\Delta f}{f_1}.
\end{equation}

By analyzing \eqref{eq_delta_f}, one can reach the same conclusions as we described for \FigRef{FIG:Beam_Squint_SEC:beam_squint_repr}. %
The beam misalignment due to beam squinting can, for example, cause reduced received power at \acp{UE}. %
This is specially critical as the number of antenna array elements increases and, therefore, beams become more and more narrow. %

\section{System Model}
\label{CHP:Beam_Squint_SEC:Sys_Mod}

In this section, we assume a generic antenna array where a filter (or beamformer) is applied. %
Then, the array beam response is presented when two different frequencies are assumed. %

\begin{figure}
	\centering
	\includetikzplot{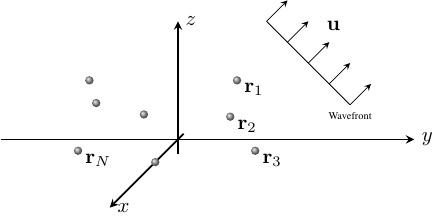}
	\caption{General antenna array representation.}
	\label{FIG:Beam_Squint_SEC:antenna_array_repr}
\end{figure}

Consider an antenna array with $N$ elements placed at positions $\vtR_{i}$, for $n \in \{1, \cdots, N\}$, as shown in \FigRef{FIG:Beam_Squint_SEC:antenna_array_repr}. %
Each element of the array has a radiation pattern given by $a(\vtU)$,  with $\vtU$ denoting the direction unit vector pointing towards the azimuth $\theta \in (-180^{\circ}, 180^{\circ})$ and zenith $\phi \in (0^{\circ}, 180^{\circ})$, i.e.,
\begin{equation}
\vtU = \Transp{\begin{bmatrix} \sin(\phi)\cos(\theta) &  \sin(\phi)\sin(\theta) &  \cos(\phi) \end{bmatrix}}.
\label{CHP:Beam_Squint_EQ:unit_vector}
\end{equation}
Furthermore, consider that a planar wavefront departs toward the direction $\vtU$ at frequency $f_c = f_1$.
In this case, the array response can be written as
\begin{equation}
\label{CHP:Beam_Squint_EQ:array_response}
\vtV(\vtU) =a(\vtU) \cdot \Transp{\begin{bmatrix}
	\psi_1 &
	\psi_2 &
	\cdots &
	\psi_N
	\end{bmatrix}},
\end{equation}
where $\psi_n = \exp \left(j \frac{2\pi f_1}{c} \Transp{\vtU} \vtR_{n}\right)$ for $n \in \{1, \cdots, N\}$.

In order to shape and steer the beam toward an arbitrary direction $\vtU_t$, the transmitter must apply a filter (or beamformer), which consists of a set of weighting factors applied on each element.
A general filter can be defined as
\begin{equation}
	\setlength\arraycolsep{3pt}
	\label{CHP:Beam_Squint_EQ:filter}
	\vtW_1 =\Transp{\begin{bmatrix} g_{1} \exp (j \omega_{1}) & g_{2} \exp (j \omega_{2}) & \cdots & g_{N} \exp (j \omega_{N})  \end{bmatrix}},
\end{equation}
where $g_n$ and $\omega_n$, for $n \in \{1, \cdots, N\}$, represent the magnitude and phase of each weight, respectively.

The beam pattern of the array combined with the filter is given by
\begin{align}
\label{CHP:Beam_Squint_EQ:beam_pattern}
B_1(\vtU) &= \Transp{\vtW}_1 \vtV(\vtU) \nonumber \\
&= a(\vtU) \sum_{n = 1}^{N} g_{n} \exp \left(j \frac{2\pi f_1}{c} \Transp{\vtU} \vtR_{n} + j \omega_{n}\right).
\end{align}

The direction $\vtU^\star$ where the array presents its maximum gain is given by
\begin{equation}
\label{CHP:Beam_Squint_EQ:max_direction}
\vtU^\star = \ArgMax{\vtU}{\Abs{B_1(\vtU)}}.
\end{equation}

Consider that the same filter $\vtW_1$ is used to beamform the array response, but operating at a frequency $f_2 = f_1 + \Delta f$.
As briefly addressed in Section~\ref{CHP:Beam_Squint_SEC:Bea_Squ_Hig_Lev_Cha}, the frequency deviation implies at a direction change.
Replacing the frequency $f_1$ in \eqref{CHP:Beam_Squint_EQ:beam_pattern} by $f_2$, the beam pattern $B_2(\vtU)$ is given by
\begin{align}
	\label{CHP:Beam_Squint_EQ:beam_pattern_offset}
	B_2(\vtU) &= a(\vtU) \sum_{n = 1}^{N} g_{n} \exp \left(j \frac{2\pi (f_1 + \Delta f)}{c} \Transp{\vtU} \vtR_{n} + j \omega_{n}\right)  \nonumber \\
	&= a(\vtU) \sum_{n = 1}^{N} g_{n} \exp \left(j \frac{2\pi f_1}{c} \Transp{\vtU} \vtR_{n}  + j \omega_{n}\right) d_n(\vtU) ,
\end{align}
where
\begin{equation}
	d_n(\vtU) = \exp \left( j \frac{2\pi \Delta f}{c} \Transp{\vtU} \vtR_{n} \right)
\end{equation}
denotes the phase deviation at each antenna element $n$.

\section{Problem Definition and Proposed Solution}
\label{CHP:Beam_Squint_SEC:Prob_Def_Prop_Sol}

Based on the antenna model and array beam pattern presented for two different frequencies, i.e., $f_1$ and $f_2$, in Section \ref{CHP:Beam_Squint_SEC:Sys_Mod}, in this section, we firstly present the phase deviation that characterizes the beam squinting. %
In order to cope with this, we then propose a filter to be applied at frequency $f_2$ to compensate for this phase deviation. %
Then, we particularize the proposed phase compensation for different array configurations. %
Finally, we discuss the main limitations of our proposal and also present a path loss compensation that can be used together with our phase compensation proposal. %

\subsection{Phase Deviation and Compensation}

Comparing $B_1(\vtU)$ in \eqref{CHP:Beam_Squint_EQ:beam_pattern} and $B_2(\vtU)$ in \eqref{CHP:Beam_Squint_EQ:beam_pattern_offset}, note that $B_2(\vtU)$ presents a $d_n(\vtU)$ at each antenna element $n$.
In order to compensate for the deviation $d_i$, the filter $\vtW_2$ used at frequency $f_2$ should be properly designed as
\begin{equation}
	\label{CHP:Beam_Squint_EQ:filter-offset}
	\vtW_2 =\vtW_1 \odot \vtC,
\end{equation}
where the operator $\odot$ denotes the Hadamard product and $\vtC$ is the compensation vector designed to cancel the effect of the deviation. %

Ideally, to completely cancel the effect of the deviation, each element $c_n$ of $\vtC$ associated with the $n$-th antenna element should be equal to $d^*_n(\vtU)$, i.e., $c_n(\vtU) = d^*_n(\vtU)$.
However, this ideal canceling is impossible, since it would require that the filter $\vtW_2$ was adapted at each wavefront direction. %
Besides that, the filter $\vtW_1$ is normally designed to be used when transmitting (or receiving) data to or from another device that is placed in a certain direction $\vtU^\star$ or in its vicinity.
This direction is determined, e.g., by the maximum array response as in \eqref{CHP:Beam_Squint_EQ:max_direction} and its vicinity depends on the beam width.

Considering the prior discussion and the assumption of knowing $\vtU^\star$ obtained in \eqref{CHP:Beam_Squint_EQ:max_direction}, we propose to define and apply a compensation vector $\vtC$ given by
\begin{equation}
	\label{CHP:Beam_Squint_EQ:compensation}
	\vtC =\Transp{\begin{bmatrix}  \psi_{1}^{\star} &  \psi_{2}^{\star} & \cdots &  \psi_{N}^{\star} \end{bmatrix}}
\end{equation}
where $\psi_{n}^{\star} =  \exp \left(- j \frac{2\pi \Delta f}{c} \Transp{(\vtU^\star)} \vtR_{n} \right) $, for $n \in \{1, \cdots, N\}$, 
to cope with the beam misalignment suffered by a beam designed to operate at frequency $f_1$ but operating at a frequency $f_2 = f_1 + \Delta f$.

This compensation method has the following advantages:
\begin{enumerate}
	\item The direction $\vtU^\star$ can be calculated once for each antenna array model, and does not depend on the antenna array placement or mechanical tilts, as long as the referential remains the same (e.g., antenna boresight);
	\item Since the compensation vector $\vtC$ only applies phase shifts, it can be used with analog, digital and hybrid beamformers;
	\item If the device uses a codebook at $f_1$, another codebook to be used at $f_2$ can be computed once and stored or even embedded at the devices during manufacturing.
\end{enumerate}

Notice that due to the usage of massive \ac{MIMO} antenna arrays in \ac{5G} and beyond, the width of the beams in these systems can be very narrow and even a small angular misalignment might have a considerable impact on the system performance. This reinforces the importance of the compensation method proposed in our work.

Besides that, in beam-based \ac{5G} and next generation systems, beam directions are often constrained to a limited number of discrete angular values, namely these of the beamforming codebook. Thus, since usually the angular positions of the \acp{UE} are considered to be the same as that of the beam selected for them for transmission and/or reception, this relatively coarse angular resolution can lead to additional angular deviations with respect to the actual angular position of the \acp{UE}. In other words, inherently the angular position of the \acp{UE} might not be known as precisely as it is assumed here, which may increase \deleted{any }harming effect\added{s} caused by angular misalignment induced by beam squinting. %
These facts strengthen even more the relevance of \replaced{our}{the herein} proposed method, which mitigates such harming effects. %

\subsection{Phase Shift Compensation for an \acs{ULA}}
\label{CHP:Beam_Squint_SEC:Prob_Def_Prop_Sol:ULA}

\begin{figure}
	\centering
	\includetikzplot{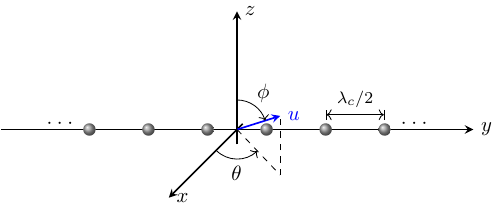}
	\caption{Representation of an \acs{ULA} placed along the $x$-axis.}
	\label{FIG:Beam_Squint_SEC:beam_squint_repr:ula_repr}
\end{figure}

Consider an \ac{ULA} with $N$ elements placed along the $y$-axis, spaced by $\frac{\lambda_c}{2}$, where $\lambda_c$ is the wavelength associated to frequency $f_c = f_1$. %
Moreover, consider that the array boresight is pointing towards the positive $x$-axis, i.e., boresight at $\theta = 0^{\text{\textdegree}}$, as depicted in \FigRef{FIG:Beam_Squint_SEC:beam_squint_repr:ula_repr}. %

The position of the $n$-th element of this \ac{ULA} is given by
\begin{equation}
	\label{CHP:Beam_Squint_EQ:ula_elem_position}
	\vtR_{n} = \Transp{\begin{bmatrix}
		0 &
		\left(n - \dfrac{N + 1}{2}\right)\dfrac{\lambda_c}{2} &
		0
	\end{bmatrix}}.
\end{equation}

Each element of the compensation vector $\vtC$ is given by
\begin{align}
		\label{CHP:Beam_Squint_EQ:compensation_ula_elem_init}
		c_n &= \exp \left(- j \frac{2\pi \Delta f}{c} \left(n - \frac{N + 1}{2}\right) \frac{\lambda_c}{2} \sin(\phi^\star)\sin(\theta^\star)  \right) \nonumber \\
		&= \exp \left(- j \frac{\pi \Delta f}{f_1} n \sin(\phi^\star)\sin(\theta^\star)  \right)  \times \nonumber \\
		& \hspace*{2.5ex} \exp \left(j \frac{\pi \Delta f}{f_1} \frac{N + 1}{2} \sin(\phi^\star)\sin(\theta^\star)  \right),
\end{align}
where $\theta^\star$ and $\phi^\star$ correspond to the azimuth and zenith where the \ac{ULA} achieves its maximum gain.
Notice that the second exponential does not depend on the antenna element, i.e., the same phase shift is applied to all elements.
It is well known that this shift does not affect the direction of the beam, therefore it can be ignored.

The \ac{ULA} does not have resolution along $\phi$, i.e., the beam pattern along $\phi$ is the same as the element pattern.
Moreover, the maximum gain of an antenna element is usually designed to point towards the boresight of the array, implying at $\phi^\star = 90^\circ$ for every adopted filter. %
Therefore, the $i$-th element of the compensation vector is given by
\begin{equation}
	\label{CHP:Beam_Squint_EQ:compensation_ula_elem}
	c_n = \exp \left(- j \frac{\pi \Delta f}{f_1} n \sin(\theta^\star)  \right), \, \forall n.
\end{equation}

\begin{figure}
	\centering
	\includetikzplot{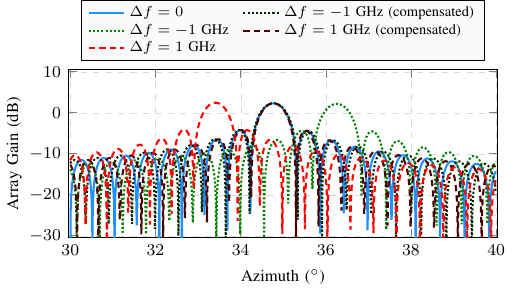}
	\caption{Illustration of beam squinting compensation for a \ac{ULA} with 256 elements and $f_c = 28\text{ GHz}$.}
	\label{FIG:Beam_Squint_SEC:beam_squint_repr:ula_comp_example}
\end{figure}

Figure \ref{FIG:Beam_Squint_SEC:beam_squint_repr:ula_comp_example} illustrates the same example presented in~\FigRef{FIG:Beam_Squint_SEC:beam_squint_repr:intermediate}, but now including two new curves. %
The new curves are related to $\Delta f = \pm 1\text{ GHz}$, while considering the proposed compensation presented in \eqref{CHP:Beam_Squint_EQ:compensation_ula_elem}. %
We observe that the main lobes of $\Delta f = 0\text{ GHz}$ and $\Delta f= \pm 1\text{ GHz}$ (compensated) match, which means that our proposal is able to mitigate the beam squinting even for a high value of $\Delta f$. %
Assuming that a given \ac{UE} is approximately aligned in the direction with azimuth $34.7^{\circ}$, at the original carrier frequency, i.e., $\Delta f = 0$, this \ac{UE} experiences maximum array gain (peak value of the solid-blue curve). %
However, if the same precoder or transmit filter is applied at a different frequency with $\Delta f = 1\text{ GHz}$, the maximum array gain is pointing towards azimuth direction of approximately $33.6^{\circ}$ (dashed-red curve). %
Thus, the aforementioned \ac{UE} would now experience an array gain much lower than the original designed one. %
The compensation is capable of keeping the maximum array gain towards the original \ac{UE} direction even at a different frequency. %
Although our compensation method focuses on aligning the main lobe of the radiation patterns for $\Delta f = 0\text{ GHz}$ and $\Delta f \neq 0\text{ GHz}$, an added benefit is that nulls near the main lobe also match for both operating frequencies as it can be seen in ~\FigRef{FIG:Beam_Squint_SEC:beam_squint_repr:ula_comp_example}. %
Thus, our method not only guarantees that the maximum gain towards a specific receiver is kept, but also avoids interference to other receivers resulting from beam squint effect. %

\subsection{Phase Shift Compensation for a \acs{URA}}
\label{CHP:Beam_Squint_SEC:Prob_Def_Prop_Sol:URA}

\begin{figure}
	\centering
	\includetikzplot{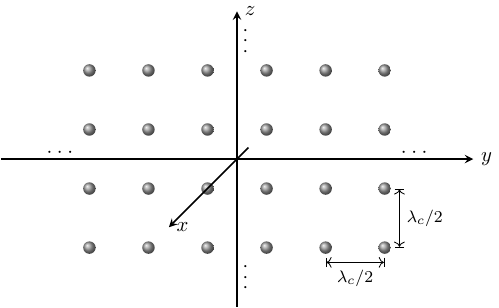}
	\caption{Representation of an \acs{URA} placed along the yz-plane, pointing towards the $x$-axis.}
	\label{FIG:Beam_Squint_SEC:beam_squint_repr:ura_repr}
\end{figure}

Consider an \ac{URA} with $N_{r} \times N_{c}$ elements placed along the $yz$-plane, spaced by $\frac{\lambda_c}{2}$. Moreover, the array boresight is pointing towards the positive $x$-axis, as depicted in \FigRef{FIG:Beam_Squint_SEC:beam_squint_repr:ura_repr}.

Following the same steps as for the \ac{ULA}, the position of the $n$-th element of this \ac{ULA} is given by
\begin{equation}
	\label{CHP:Beam_Squint_EQ:ura_elem_position}
	\vtR_{n} = \Transp{\begin{bmatrix}
		0 &
		\left(n_r - \dfrac{N_r + 1}{2}\right)\dfrac{\lambda_c}{2} &
		\left(n_c - \dfrac{N_c + 1}{2}\right)\dfrac{\lambda_c}{2}
	\end{bmatrix}},
\end{equation}
where $n_r \in \{1, \cdots, N_r\}$ and $n_c \in \{1, \cdots, N_c\}$ denote the index of the row ($z$-axis) and column ($y$-axis) of the \ac{URA}, respectively.

Considering the element positioning in \eqref{CHP:Beam_Squint_EQ:ura_elem_position} and removing the antenna independent phase shift, the same as in \eqref{CHP:Beam_Squint_EQ:compensation_ula_elem_init} and \eqref{CHP:Beam_Squint_EQ:compensation_ula_elem} for the \ac{ULA}, each element of the compensation vector $\vtC$ is given by %
\begin{equation}
	\label{CHP:Beam_Squint_EQ:compensation_ura_elem}
	c_n = \exp \left(- j \frac{\pi \Delta f}{f_1} \left(n_r \sin(\phi^\star)\sin(\theta^\star)  + n_c  \cos(\phi^\star)\right)\right).
\end{equation} %

\begin{figure}
	\subfloat[Azimuth.]{%
		\centering
		\includetikzplot{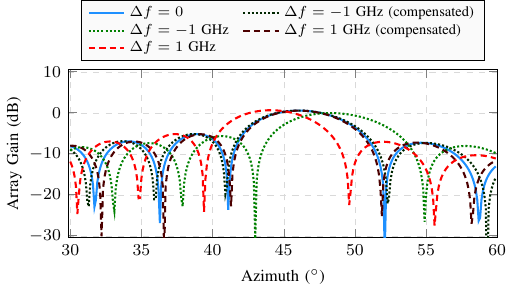}
		\label{FIG:Beam_Squint_SEC:beam_squint_repr:ura_comp_az_example}
	}
	
	\subfloat[Zenith.]{%
		\centering
		\includetikzplot{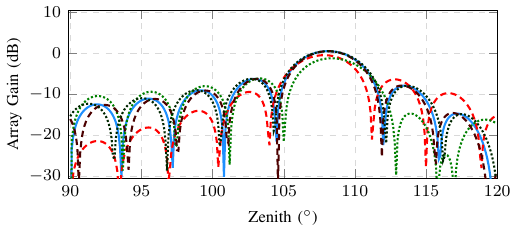}
		\label{FIG:Beam_Squint_SEC:beam_squint_repr:ura_comp_zen_example}
	}
	\caption{{Illustration of beam squinting compensation for a \ac{URA} $32\times32$ and $f_c = 28\text{ GHz}$.}}
	\label{FIG:Beam_Squint_SEC:beam_squint_repr:ura_comp_example}
\end{figure}

Figure \ref{FIG:Beam_Squint_SEC:beam_squint_repr:ura_comp_example} illustrates the proposed beam squinting compensation for a $32\times32$ \ac{URA}. %
This figure presents the gain of this \ac{URA} for five different cases. %
Three different values of $\Delta f$ are evaluated, i.e., $\Delta f = 0\text{ GHz}$ and $\Delta f = \pm 1 \text{ GHz}$. %
For $\Delta f= \pm 1 \text{ GHz}$, two possibilities are considered, i.e., with and without the proposed beam squinting compensation. %
Comparing these five cases we can conclude that our proposal is able to compensate for the beam squinting that occurs for $\Delta f = \pm 1 \text{ GHz}$. %
The advantage in terms of interference avoidance as discussed for~\FigRef{FIG:Beam_Squint_SEC:beam_squint_repr:ula_comp_example} can also be observed in this case. %

\subsection{Limitations of the Proposed Phase Compensation Method}

As observed in Figs. \ref{FIG:Beam_Squint_SEC:beam_squint_repr:ula_comp_example} and \ref{FIG:Beam_Squint_SEC:beam_squint_repr:ura_comp_example}, the proposed solution successfully compensates for the beam squinting effect caused by frequency displacement.
However, if the filter $\vtW$ yields an array response with main lobes pointing towards more than one direction, the solution proposed in the previous section may fail. %
This case may occur when the beam is steered close to the array endfire.

\begin{figure}
	\centering
	\includetikzplot{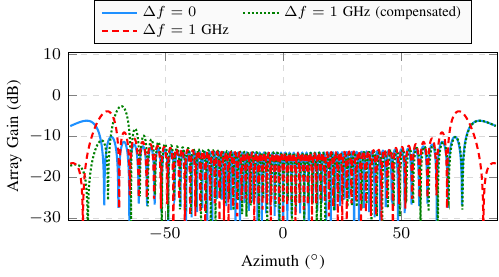}
	\caption{Array gain of a \ac{ULA} with $64$ elements with beam steering close to the array endfire.}
	\label{FIG:Beam_Squint_SEC:beam_squint_repr:ula_endfire}
\end{figure}

As an example, consider a ULA with 64 elements disposed along the $y$-axis and a filter $\vtW$ equal to the column of the DFT codebook that steers the beam closest to the array endfire.
When applying our proposed compensation, the array response is compensated in only one direction, as depicted in \FigRef{FIG:Beam_Squint_SEC:beam_squint_repr:ula_endfire}. %

Notice, in \FigRef{FIG:Beam_Squint_SEC:beam_squint_repr:ula_endfire}, that the chosen filter fits to serve devices placed at an azimuth direction around $[-90^\circ, -80^\circ]$ or $[80^\circ, 90^\circ]$.
As already discussed, the proposed solution needs prior knowledge of which direction $\vtW$ steers the array to in order to properly compensate it.
In this example, only the positive azimuth direction is compensated, which yields to an inappropriate antenna gain if the served device is placed at the negative azimuth.

In order to address this issue, when the value $\vtW$ leads to more than one beam direction, the transmitter should make use of the device's direction, in order to properly compensate for the beam squinting effect.
The device's direction can be estimated using, for example, the beam tracking procedure, or the \ac{SSB} measurements.
Therefore, considering the case where $\vtW$ beamforms the array towards $D$ different directions $\vtU^\star_i$ for $i \in \{1,\ldots,D\}$, and the served device's estimated direction is $\tilde{\vtU}$, the proposed solution should compensate for the beam squinting towards the direction $\vtU^\star = \vtU^\star_{i^{\dagger}}$
\begin{equation}
\label{CHP:Beam_Squint_EQ:max_direction_several}
i^{\dagger} = \ArgMin{i}{\Norm{\vtU^\star_i - \tilde{\vtU}}}.
\end{equation}

On the other hand, the filters are normally designed to address devices upon a single direction, and are taken from a predefined set, like the aforementioned example, where $\vtW$ was taken from the \ac{DFT} codebook. %
Moreover, the array is normally designed to irradiate most of its energy to its front. %
Therefore, in such use cases, the only situation where $\vtW$ beamforms the array response to more than one direction is when $\vtW$ steers the response close to the array endfires, as depicted in \FigRef{FIG:Beam_Squint_SEC:beam_squint_repr:ula_endfire}. %
In this situation the strategy proposed in \eqref{CHP:Beam_Squint_EQ:max_direction_several} can be quite simplified. %
When the array is an \ac{ULA} and the \ac{DFT} codebook is employed, just one filter (one codebook entry) has ambiguous directions. %
In order to disambiguate it, the transmitter should estimate whether the azimuth of the served device with respect to the array boresight is greater than 0. %
If an \ac{URA} is considered, some filters of the codebook need disambiguation, due to azimuth and zenith endfires. %
In this case, besides the azimuth, the transmitter must estimate whether the zenith is greater than $90^\circ$. %

\subsection{Path Loss Compensation}

Besides the phase compensation discussed in this section, the use of different frequencies can also impact on long-term channel characteristics, e.g., path loss. %
Thus, assume a generic path loss model (in dB)

\begin{equation} \label{CHP:Beam_Squint_EQ:Path_loss}
L_{k} = a + b\cdot\log_{10}\left(d_{3D}\right )+c\cdot\log_{10}\left(f_{k}\right ),
\end{equation}
\noindent where $a$, $b$ and $c$ are constants of the model, $d_{3D}$ is the distance between the transmitter and receiver, and $f_{k}$ is the considered frequency. %
By assuming two different frequencies, $f_1$ and $f_2$ where $f_2 = f_1 + \Delta f$ and $f_2 > f_1$, we can show that the positive difference in path loss is
\begin{equation} \label{CHP:Beam_Squint_EQ:PL_diff}
L_{2}-L_{1} = c\cdot\log10\left(1+\frac{\Delta f}{f_{1}}\right).
\end{equation}

For $f_{1}$ in the order of tens of GHz (\ac{mmWave}) and $\Delta f$ in the order of hundreds of MHz, $L_{2}-L_{1}$ will be small. %
For example, considering $f_{1}=\SI{28}{GHz}$, for $L_{2}-L_{1} \geq \SI{0.3}{dB}$, then $\Delta f\geq \SI{1}{GHz}$. %
Thus, for $f_{k}$ in \ac{mmWave}, as in \ac{5G}, we do not expect an important variation of the path loss in a bandwidth
of a few hundreds of MHz. %

\section{Performance Evaluation: Case Study with \acs{5G}/\acs{NR} and \acsp{NCR}}
\label{CHP:Beam_Squint_SEC:Perf_Eval}

Although our proposed beam compensation method applies to general multiple-antenna transmitters, e.g., \acp{BS}, \acp{RIS} and \acp{NCR}, in this section, we present the simulation results for \ac{NCR}-assisted networks to evaluate the impact of our beam squinting compensation proposal in the \ac{NCR} subband operation.\footnote{While we present the setup for \ac{NCR}-assisted networks, the same approach is well-applicable for \ac{RIS}-assisted networks.} %
\added{Here, we consider system-level analysis where the effect on beam squinting on both the useful and interfering signals can be well studied.} %
In our performance evaluation, we consider the downlink direction and that the beam squinting compensation method is applied at both \ac{gNB} and \ac{NCR} nodes. %
\acp{NCR} were chosen in our analysis \replaced{because, different from \acp{RIS}, have been standardized by 3GPP}{since they are in a more advanced standardization stage at \ac{3GPP},} and have shown better performance when compared to \acp{RIS}~\cite{Guo2022}. %
Another motivation for this case study is that \Acp{NCR} may be deployed in crowded areas where their distance to the \acp{UE} may be not high and the \acp{UE} are not necessarily aligned to the array boresight direction and they can be served by beams close to the antenna array endfire. %
Thus, depending on the \ac{NCR} deployment, it is expected for the beam squinting phenomenon to have an important impact on \ac{NCR} subband operation. %

The details concerning the considered simulation modeling are presented in \SecRef{CHP:Beam_Squint_SEC:Sim_Param} and the results are discussed in \SecRef{CHP:Beam_Squint_SEC:Sim_Results}. %

\subsection{Simulation Scenario, \acs{NCR} Modeling and Parameters} \label{CHP:Beam_Squint_SEC:Sim_Param}

\begin{figure}[!t]
	\centering
	\includegraphics[width=0.7\columnwidth]{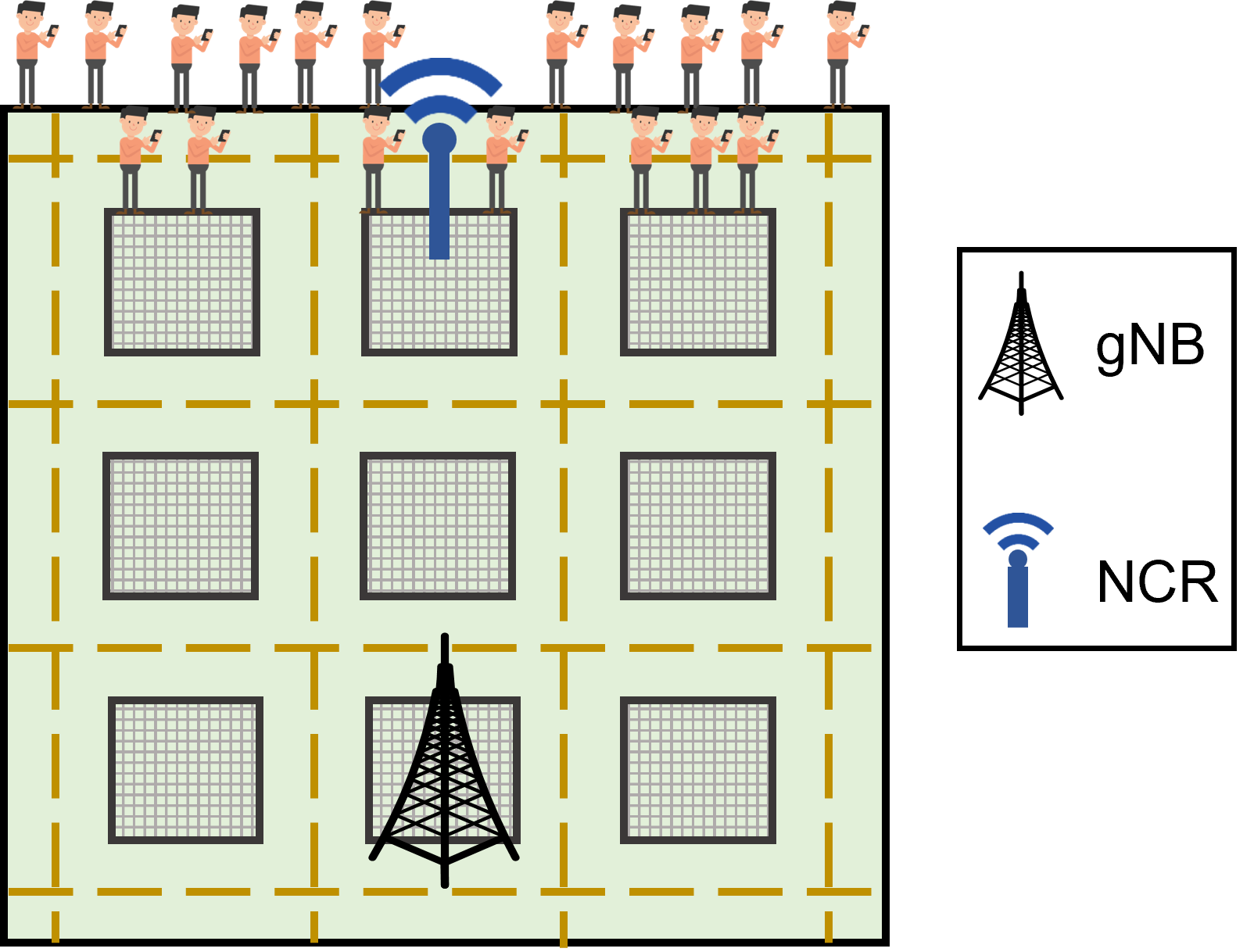}
	\caption{Considered scenario.}
	\label{CHP:Beam_Squint_FIG:scenario}
\end{figure}

Figure \ref{CHP:Beam_Squint_FIG:scenario} illustrates the adopted scenario\footnote{While the simulations try to mimic the Rel-18 NCR specifications, the considered setup is not necessarily aligned with Ericsson point of view about NCRs-assisted networks.}. %
A simplified version of the Madrid grid~\cite{METIS:D6.1:2013} is considered. %
In this scenario, there are $9$~square blocks, with dimension of \SI{120}{m}~$\times$~\SI{120}{m} surrounded by \SI{3}{\meter}~wide sidewalks and separated of each other by \SI{14}{m}~wide streets. %
The \acp{UE} are randomly placed in the sidewalks of a crowded street. %
They are allowed to cross the street only in the intersections. %
The closest \ac{gNB} is deployed some blocks away. %
An \ac{NCR}, controlled by the \ac{gNB}, is deployed in the street where the \acp{UE} are located in order to improve the system coverage in that area. %

Regarding the \ac{NCR}, at each \ac{TTI} $t$, it applies a fixed gain of 60 dB  to the signal received at each \ac{RB}. If the signal output power of the \ac{NCR} with fixed gain is higher than the maximum \ac{NCR} output power, the gain $g_{k}$ at \ac{RB} $k$ will be given as
\begin{equation}\label{CHP:Beam_Squint_EQ:NCR_gain}
g_{k}(t) = \frac{p^{\text{NCR}}_{k}}{\sigma_{k}^{2}+\gamma_{k}(t)\cdot p_{\tx,k}(t)},
\end{equation}
where $\sigma_{k}$ is the noise power at  the bandwidth of \ac{RB} $k$; $\gamma_{k}(t) = |\textbf{d}_{k}(t)
\textbf{Q}_{k}(t) \textbf{w}_{k}(t)|^{2}$ denotes the combined effect of the channel $\textbf{Q}_{k}$ after the
transmission filter at the \ac{gNB}, $\textbf{w}_{k}$, and the reception filter at \ac{NCR}, $\textbf{d}_{k}$, applied to a signal transmitted with transmit power $p_{\tx,k}$; and $p^{NCR}_{k}$ is the output power of the \ac{NCR}, which is constant, since it is considered \ac{EPA} among the \acp{RB}. %

All \deleted{the }links between the nodes in the system, i.e., \acp{UE}, \ac{NCR} and \ac{gNB}, are modeled according to the \ac{3GPP} channel model specifications~\cite{3gpp.38.901_new,Pessoa2019}. %
It is a spatially and \replaced{temporally}{time} consistent model that considers a distance-dependent path-loss, a lognormal shadowing component and small-scale fading. %

Concerning the \acp{UE}' signal strength measurements at $f_1$, the \ac{gNB} periodically performs beam sweeping with \acp{SSB} and \ac{CSI-RS}~\cite{Monteiro2019}. %
Some of the \acp{SSB}/\acp{CSI-RS} are amplified and forwarded by the \ac{NCR}, while others arrive directly from the \ac{gNB} to the \acp{UE}. %
For each received \ac{SSB}/\ac{CSI-RS}, a \ac{UE} measures its \ac{RSRP}, which is the linear average over the power contributions (in Watts) of the resource element confined within a transmitted \ac{SSB}/\ac{CSI-RS}~\cite{3gpp.38.215}. %
The measured \acp{RSRP} are then reported to the \ac{gNB}. %
Based on the received measurement reports, the \ac{gNB} chooses whether the \ac{UE} should be served directly or through the \ac{NCR}. %
Similarly, the \ac{NCR} also measures and reports to the \ac{gNB} the \ac{RSRP} of the received \acp{SSB}/\acp{CSI-RS} in order to allow the \ac{gNB} to perform link adaptation of the control and backhaul links (see Fig. \ref{CHP:Beam_Squint_FIG:proposal}). %

The links between \ac{gNB} and \ac{NCR}, i.e., control and backhaul, are stable and planned during network deployment. %
On the other hand, the links between \acp{UE}-\ac{gNB} and \acp{UE}-\ac{NCR} vary more frequently due to the \acp{UE} mobility. %
Thus, the beam sweeping of beams related to control and backhaul links are performed less frequently than the beam sweeping of beams related to access links. %

The system parameters are aligned with \ac{NR} \ac{3GPP} specifications series~$38$ Release~$17$. %
The simulations are conducted at \SI{28}{\GHz}. %
Four different values of $\Delta f$ are considered: \SI{0}{MHz}, \SI{100}{MHz}, \SI{500}{MHz} and \SI{1}{GHz}. 
An \ac{RB}, i.e., the minimum scheduling unit in the frequency domain, consists of $12$~consecutive subcarriers spaced \SI{60}{kHz} of each other. %
Thus, $66$~\acp{RB} are considered. %
Moreover, we consider an \ac{AWGN} power per subcarrier of \SI{-174}{dBm}, noise figure of \SI{9}{dB} and shadow standard deviation of \SI{4}{dB}. %
In the time domain, a slot is the minimum scheduling unit, consisting of $14$~\ac{OFDM} symbols with total time duration of \SI{0.25}{\ms}. %

Regarding resource scheduling, in the frequency domain, the \ac{RR} scheduler is adopted to schedule the \acp{RB}. %
The \ac{RR} iteratively allocated the \acp{RB}, scheduling in a given \ac{RB} the \ac{UE} bearer waiting the longest time in the queue. %
The \ac{RR} is chosen since it is a well-known scheduler enabling an interested reader to reproduce our performance evaluation. %
Besides, our main objective is not to find the scheduler that optimizes the system behavior, but rather compare the different configurations under the same conditions, i.e., using the same scheduler. %

The \ac{CQI}/\ac{MCS} mapping curves standardized in~\cite{3gpp.38.214} are used for link adaptation with a target \ac{BLER} of \SI{10}{\%}. %
It is also considered an outer loop strategy to avoid the increas\replaced{e}{ing} of the \ac{BLER}. %
According to this strategy, when a transmission error occurs, the estimated \ac{SINR} used for the \ac{CQI}/\ac{MCS} mapping in the link adaptation is subtracted by a back-off value of \SI{1}{dB}. %
On the other hand, when a transmission occurs without error, the estimated \ac{SINR} has its value added by \SI{0.1}{dB}. %

In the considered scenario, there are $72$~\acp{UE} with traffic modeled as \ac{CBR} flows with packet size equal to 4,096~bits and packet-inter-arrival time equal to four slots. %
\TabRef{CHP:Beam_Squint_TABLE:Entities-characteristics} presents other parameters considered in the simulations. %

In the following, the computational results are presented.
For each value of $\Delta f$, three possibilities are evaluated:
\begin{itemize}
	\item \ac{CP} transmitted at \SI{28}{GHz} and \ac{DP} transmitted at $($\SI{28}{GHz}+$\Delta f)$ without compensation,
	\item \ac{CP} transmitted at \SI{28}{GHz} and \ac{DP} transmitted at $($\SI{28}{GHz}+$\Delta f)$ with compensation, and,
	\item Both \ac{CP} and \ac{DP} transmitted at $($\SI{28}{GHz}+$\Delta f)$,
\end{itemize}
where, as already mentioned, \ac{CP} performs the transmission of signaling related to measurements used for \ac{RRM}, e.g., link adaptation, and \ac{DP} performs \ac{UE} data transmission. %

Finally, three \ac{gNB} antenna array configurations are evaluated: \ac{ULA} with $64, 128 \text{ and } 256$~antenna elements. %
\added{It is important to highlight that \ac{3GPP} plans to study antenna arrays with more than 64 elements in Rel. 19.} %
Three \acp{KPI} are considered: the \ac{SINR}, the throughput and the \ac{MCS} usage, all of them in the \ac{DL}. %

\begin{table}[!t]
	\centering
	\caption{Entities characteristics.}
	\label{CHP:Beam_Squint_TABLE:Entities-characteristics}
	\begin{tabularx}{\columnwidth}{lXXX}
		\toprule
		\textbf{Parameter} & \textbf{\ac{gNB}} & \textbf{\ac{NCR}} & \textbf{\ac{UE}} \\
		\midrule
		Height & \SI{25}{\meter} & \SI{10}{\meter} & \SI{1.5}{\meter} \\
		Transmit power & \SI{35}{\decibel m} & \SI{33}{\decibel m} & \SI{24}{\decibel m} \\
		Antenna element pattern & \ac{3GPP} 3D~\cite{3gpp.38.901_new} & \ac{3GPP} 3D~\cite{3gpp.38.901_new} & Omni \\
		Max. antenna element gain & \SI{8}{\decibel i} & \SI{8}{\decibel i} & \SI{0}{\decibel i} \\
		Speed & \SI{0}{km/h} & \SI{0}{km/h}  & \SI{3}{km/h} \\
		\bottomrule
	\end{tabularx}
\end{table}

\subsection{Simulation Results} \label{CHP:Beam_Squint_SEC:Sim_Results}

\begin{figure}
	\centering
	\subfloat[ULA 64.]{%
		\centering
		\includetikzplot{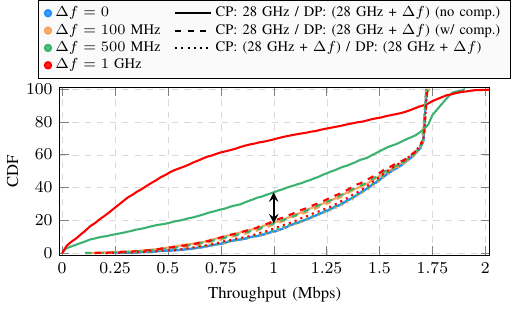}
		\label{FIG:Beam_Squint_SEC:thoughput_down:ula64}
	}
	
	\subfloat[ULA 128.]{%
		\centering
		\includetikzplot{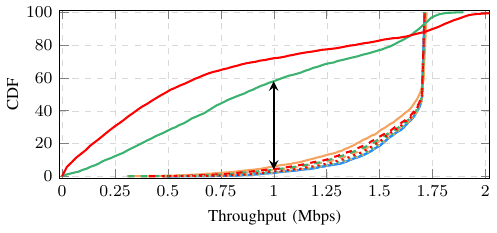}
		\label{FIG:Beam_Squint_SEC:thoughput_down:ula128}
	}
	
	\subfloat[ULA 256.]{%
		\centering
		\includetikzplot{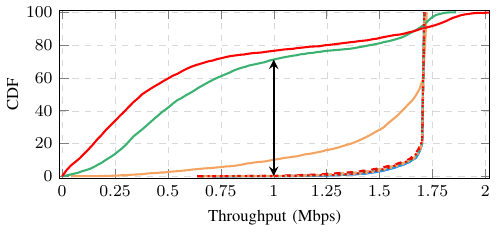}
		\label{FIG:Beam_Squint_SEC:thoughput_down:ula256}
	}
	\caption{\acs{CDF} of \acs{UE} \acs{DL} throughput.}
	\label{FIG:Beam_Squint_SEC:thoughput_down}
\end{figure}

Figure \ref{FIG:Beam_Squint_SEC:thoughput_down} presents the \ac{CDF} of the \acp{UE}' \ac{DL} throughput. %
First, we observe that when \ac{CP} and \ac{DP} are transmitted at the same frequency, i.e., blue solid curve ($\Delta f=0$) and dotted curves, the \acp{UE} experience similar throughput. %
However, comparing these cases, i.e., \ac{CP} and \ac{DP} transmitted at the same frequency (blue solid curve and dotted curves) with the cases where \ac{CP} and \ac{DP} are transmitted at different frequencies without compensation (solid curves), we observe that the \acp{UE} throughput decreases. %
The impact of beam squinting increases for higher values of $\Delta f$, as expected. %

Comparing the cases with and without our compensation proposal, i.e., solid and dashed curves, respectively, we observe that with our proposal  the \acp{UE}' throughput is similar to the cases when \ac{CP} and \ac{DP} were transmitted at the same frequency. %
Then, we can conclude that our proposed method is able to overcome the beam squinting issue. %
Furthermore, comparing Figures~\ref{FIG:Beam_Squint_SEC:thoughput_down:ula64}, \ref{FIG:Beam_Squint_SEC:thoughput_down:ula128} and \ref{FIG:Beam_Squint_SEC:thoughput_down:ula256}, we observe that the beam squinting problem becomes more dominant (solid curves moved to the left) when the number of antenna elements increases, i.e., when we consider narrower beams. %
However, in all cases, our proposal is well able to compensate for the beam squinting. %
In these figures, we can see that the percentage of \acp{UE} with throughput lower than 1 Mbps assuming $\Delta f = 500 \text{MHz}$ and no compensation are equal to 37\%, 57\% and 71\% whereas these values decrease to 18\%, 4\% and 0.2\% when our proposed compensation method is applied for \ac{ULA} with 64, 128 and 256 elements, respectively. %

\begin{figure}
	\centering
	\subfloat[ULA 64.]{%
		\centering
		\includetikzplot{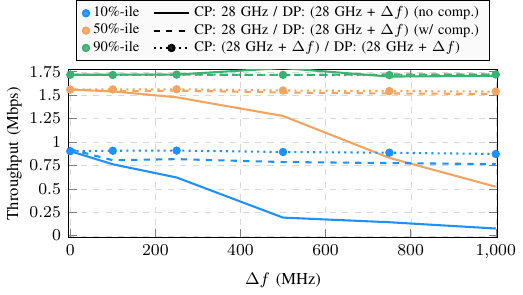}
		\label{FIG:Beam_Squint_SEC:thoughput_by_offset_down:ula64}
	}
	
	\subfloat[ULA 128.]{%
		\centering
		\includetikzplot{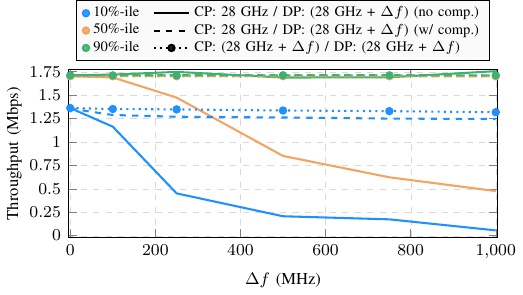}
		\label{FIG:Beam_Squint_SEC:thoughput_by_offset_down:ula128}
	}
	
	\subfloat[ULA 256.]{%
		\centering
		\includetikzplot{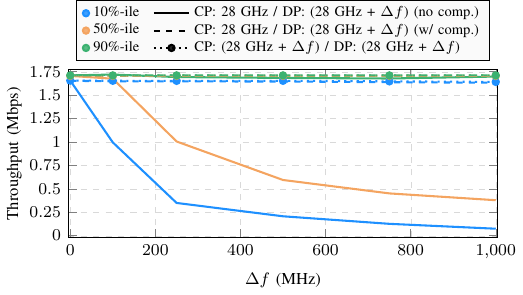}
		\label{FIG:Beam_Squint_SEC:thoughput_by_offset_down:ula256}
	}
	\caption{10, 50 and 90\%-ile of the \acs{UE} \acs{DL} throughput.}
	\label{FIG:Beam_Squint_SEC:thoughput_by_offset_down}
\end{figure}

Figure \ref{FIG:Beam_Squint_SEC:thoughput_by_offset_down} presents another perspective of the one shown in Figure \ref{FIG:Beam_Squint_SEC:thoughput_down}. %
In this figure, we show the throughput versus the frequency deviation assuming the cases without compensation, with compensation and the one where \ac{CP} and \ac{DP} are transmitted at the same frequency (no beam squinting). %
Figures \ref{FIG:Beam_Squint_SEC:thoughput_by_offset_down:ula64}, \ref{FIG:Beam_Squint_SEC:thoughput_by_offset_down:ula128} and \ref{FIG:Beam_Squint_SEC:thoughput_by_offset_down:ula256} are for \acp{ULA} with 64, 128 and 256 antenna elements, respectively. %
Firstly, we can see that the cases without beam squinting, without compensation and with compensation presents a similar performance at the $90\%\text{-ile}$, which indicates that beam squint is not an issue for the \acp{UE} with high throughput, e.g., cell center \acp{UE}. %
However, beam squinting imposes an increasing performance loss as the frequency deviation increases for \acp{UE} in medium ($50\%\text{-ile}$) and poor ($10\%\text{-ile}$) channel conditions. %
Furthermore, we \replaced{observe}{can see} that the case with \ac{ULA} 256 is the case where there is the highest performance loss when frequency deviation increases for medium and low percentiles illustrating the sensitivity of beam squinting for large antenna arrays. %
Our compensation method shows an almost constant performance with respect to the frequency deviation and very close to the case where \ac{CP} and \ac{DP} are transmitted at the same frequency. %
This shows the advantage of employing our method that is capable of keeping high throughput by compensating for the beam squinting effect especially for \acp{UE} in medium and poor channel conditions. %
\added{This is particularly important because the main motivation for \acp{NCR} is to boost the performance of weak \acp{UE}}. %

%
%

\begin{figure}
	\centering
	\subfloat[ULA 64.]{%
		\centering
		\includetikzplot{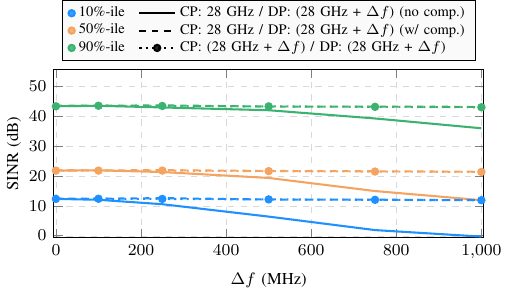}
		\label{FIG:Beam_Squint_SEC:sinr_by_offset_down:ula64}
	}
	
	\subfloat[ULA 128.]{%
		\centering
		\includetikzplot{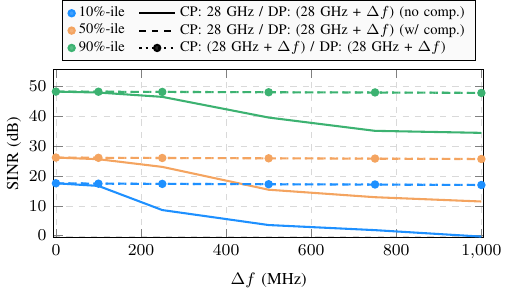}
		\label{FIG:Beam_Squint_SEC:sinr_by_offset_down:ula128}
	}
	
	\subfloat[ULA 256.]{%
		\centering
		\includetikzplot{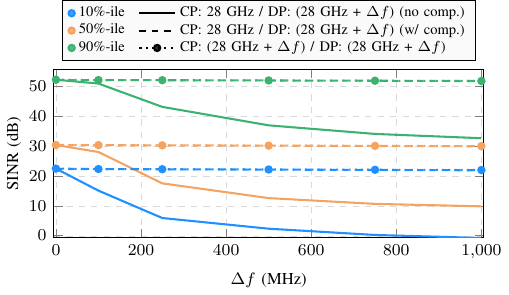}
		\label{FIG:Beam_Squint_SEC:sinr_by_offset_down:ula256}
	}
	\caption{10, 50 and 90\%-ile of the \acs{UE} \acs{DL} \acs{SINR}.}
	\label{FIG:Beam_Squint_SEC:sinr_by_offset_down}
\end{figure}

In Fig. \ref{FIG:Beam_Squint_SEC:sinr_by_offset_down}, we present the \ac{SINR} versus the frequency deviation assuming the cases without compensation, with compensation and the one where \ac{CP} and \ac{DP} are transmitted at the same frequency (no beam squinting). %
Figures \ref{FIG:Beam_Squint_SEC:sinr_by_offset_down:ula64}, \ref{FIG:Beam_Squint_SEC:sinr_by_offset_down:ula128} and \ref{FIG:Beam_Squint_SEC:sinr_by_offset_down:ula256} are for \acp{ULA} with 64, 128 and 256 antenna elements, respectively. %
As the \acp{SINR} are mapped to data rates through the link adaptation mechanism, Figs. \ref{FIG:Beam_Squint_SEC:thoughput_by_offset_down} and \ref{FIG:Beam_Squint_SEC:sinr_by_offset_down} presents some similar aspects, but there are other important conclusions to highlight. %
The differences in \ac{SINR} among 90\%-ile, 50\%-ile and 10\%-ile are more prominent when compared with the same ones for throughput in the same percentiles in Fig. \ref{FIG:Beam_Squint_SEC:thoughput_by_offset_down}. %
The reason for that is the data rate saturation as the number of \ac{MCS} schemes is finite, i.e., although the difference in \ac{SINR} between 90\%-ile and 50\%-ile are more than 20 dB, this difference does not directly translate into a large difference in data rate. %

In Fig. \ref{FIG:Beam_Squint_SEC:sinr_by_offset_down} we can also have another view on how system performance, in terms of \ac{SINR}, degrades due to beam squinting with respect to frequency deviation and antenna array size. %
For the 90\%-ile (\acp{UE} in best channel conditions), the \ac{SINR} degradation due to beam squinting becomes more intense for frequency deviations of approximately 500 MHz, 200 MHz and 100 MHz for \ac{ULA} with 64, 128 and 256 antenna elements, respectively. %
The sensitivity of the \ac{SINR} of \acp{UE} in intermediate and worst channel conditions is higher with respect to frequency deviation compared to the 90\%-ile case, as it can be seen in orange and blue curves. %
\acp{UE} with worst channel conditions are in general the ones far away from the \ac{gNB}/\ac{NCR} and also in the endfire of the antenna arrays where beam squinting is more significant, as previously explained. %

\begin{figure}
	\centering
	\subfloat[$\Delta f = 100$ MHz.]{%
		\centering
		\includetikzplot{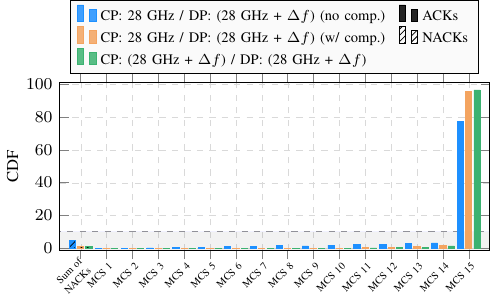}
		\label{FIG:Beam_Squint_SEC:mcs_down:ula256_100}
	}
	
	\subfloat[$\Delta f = 500$ MHz.]{%
		\centering
		\includetikzplot{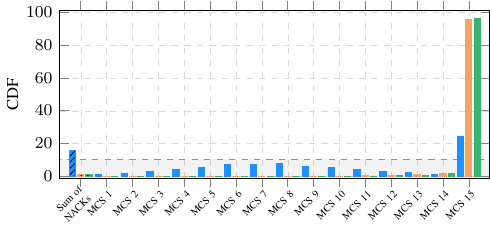}
		\label{FIG:Beam_Squint_SEC:mcs_down:ula256_500}
	}
	
	\subfloat[$\Delta f = 1$ GHz.]{%
		\centering
		\includetikzplot{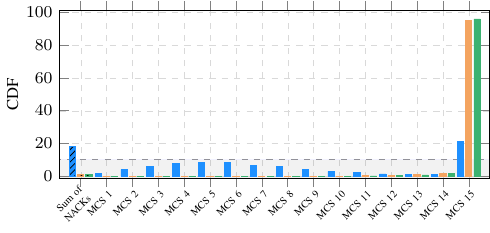}
		\label{FIG:Beam_Squint_SEC:mcs_down:ula256_1000}
	}
	\caption{Histogram of \acp{UE}' \ac{MCS} usage in \ac{DL} for different values of $\Delta f$ considering a \ac{gNB} with a \ac{ULA} with 256 antenna elements.}
	\label{FIG:Beam_Squint_SEC:mcs_down:ula256}
\end{figure}

Fig. \ref{FIG:Beam_Squint_SEC:mcs_down:ula256} presents the histogram of \acp{UE}' \ac{MCS} usage in \ac{DL} for different values of $\Delta f$ considering a \ac{gNB} with an \ac{ULA} with 256 antenna elements. %
It compares three cases: i) \ac{CP} and \ac{DP} transmitted at the same frequency $(28\text{GHz}+\Delta f)$; ii) \ac{CP} transmitted at \SI{28}{GHz} and \ac{DP}, at $(28\text{GHz}+\Delta f)$, without compensation; and iii) \ac{CP} transmitted at \SI{28}{GHz} and \ac{DP}, at $(28\text{GHz}+\Delta f)$, with compensation. %

As seen in Fig. \ref{FIG:Beam_Squint_SEC:mcs_down:ula256}, with our proposal, when transmitting \ac{CP} at \SI{28}{GHz} and \ac{DP} at $(28\text{GHz}+\Delta f)$,  the system achieves similar performance as when transmitting both \ac{CP} and \ac{DP} at the same frequency. %
On the other hand, without our proposal, the beam squinting deteriorates the system performance. %

Notice in Figs.~\ref{FIG:Beam_Squint_SEC:mcs_down:ula256_500} and~\ref{FIG:Beam_Squint_SEC:mcs_down:ula256_1000}\deleted{,} that for $\Delta f=500\text{ MHz}$ and $\Delta f=1\text{ GHz}$, respectively, when \ac{CP} is transmitted at \SI{28}{GHz} and \ac{DP} at $(28\text{ GHz}+\Delta f)$, the sum of NACKs was higher than the target \ac{BLER} of $10\%$ adopted in the link adaptation. %
This means that the beam squinting causes more transmission failures than the acceptable target. %
The higher number of NACKs is a consequence of the \ac{SINR} degradation shown in Fig. \ref{FIG:Beam_Squint_SEC:sinr_by_offset_down} induced by the beam squinting phenomenon. %
Even worse, we observe that this effect increases with the increase of $\Delta f$, which highlights the necessity of adopting beam squinting compensation when transmitting data at $f_c+\Delta f$ based on measurements performed at frequency $f_c$. %

\FloatBarrier
\section{Conclusions}
\label{CHP:Beam_Squint_SEC:Conclusion}

This paper proposed an efficient compensation method for beam squinting in repeater-assisted networks. %
From a link level point-of-view, we have shown that our proposed method is able to compensate for the beam squinting effect on the array gain in both azimuth and zenith for different values of frequency displacement and different antenna array configurations. %
As a consequence, from a system level point-of-view, we have shown through simulations that our proposal enables the \ac{NCR} subband signaling operation. %
With our proposal, when transmitting data at a frequency different from the one in which the measurements are performed, we are able to keep \acp{UE}' throughput at the same level as when the data transmission and channel measurement are performed at the same frequency. %
Furthermore, we have shown that, without a compensation, the perceived \ac{SINR}, and so \acp{UE}' throughput, can decrease substantially. %
For instance, considering an \ac{ULA} with 256 antenna elements, $f_c=28\text{GHz}$ and $\Delta f=1\text{GHz}$. %
Without compensation, the \ac{SINR} of the 10-percentile decreases by \SI{20}{dB} compared to the case when \ac{CP} and \ac{DP} operate at the same frequency. %
This is mainly due to the beam misalignment to the intended receiver as well as the increase in the interference to other receivers due to beam squinting. %
Such a degradation is fully compensated with our proposed method. %


\ifCLASSOPTIONcaptionsoff
  \newpage
\fi

\printbibliography
\end{document}

%% file: acronymsLowerCase.tex
\acro{2D}{two-dimensional}
\acro{3D}{three-dimensional}
\acro{4D}{four-dimensional}
\acro{6D}{six-dimensional}
\acro{1G}{first generation}
\acro{2G}{second generation}
\acro{3G}{third generation}
\acro{4G}{fourth generation}
\acro{5G}{fifth generation}
\acro{5GC}{5G core network}
\acro{5G-StoRM}{5G stochastic radio channel for dual mobility}
\acro{6G}{sixth generation}
\acro{3GPP}{3rd generation partnership project}
\acro{3GPP2}{3rd generation partnership project 2}
\acro{A2A}{air-to-air}
\acro{A2G}{air-to-ground}
\acro{AA}{antenna array}
\acro{AC}{admission control}
\acro{AD}{attack-decay}
\acro{AE}{antenna element}
\acro{AF}{amplify and forward}
\acro{ABS}{almost blank subframe}
\acro{ACF}{autocorrelation function}
\acro{ADSL}{asymmetric digital subscriber line}
\acro{AHW}{alternate hop-and-wait}
\acro{AI}{artificial intelligence}
\acro{AMC}{adaptive modulation and coding}
\acro{AP}{access point}
\acro{APA}{adaptive power allocation}
\acro{API}{application protocol interface}
\acro{ARQ}{automatic repeat request}
\acro{AT}{averaged time}
\acro{ARMA}{autoregressive moving average}
\acro{ASE}{average squared error}
\acro{ASC}{adaptive satisfaction control}
\acro{ATES}{adaptive throughput-based efficiency-satisfaction trade-off}
\acro{AWGN}{additive white gaussian noise}
\acro{B5G}{beyond 5G}
\acro{BAP}{backhaul adaptation protocol}
\acro{BB}{branch and bound}
\acro{BC}{branch and cut}
\acro{BD}{block diagonalization}
\acro{BER}{bit error rate}
\acro{BF}{best fit}
\acro{BL}{bit loading}
\acro{BLER}{block error rate}
\acro{BLPC-1}{bit loading with power constraint in hop 1}
\acro{BLPC-2}{bit loading with power constraint in hop 2}
\acro{BPC}{binary power control}
\acro{BPSK}{binary phase-shift keying}
\acro{BRA}{balanced random allocation}
\acro{BS}{base station}
\acro{BSP}{\acs*{BS} placement}
\acro{BSR}{buffer status report}
\acro{BoI}{band of interest}
\acro{C-link}{control link}
\acro{CAP}{combinatorial allocation problem}
\acro{CAPEX}{capital expenditure}
\acro{CB}{contextual bandit}
\acro{CBF}{coordinated beamforming}
\acro{CBR}{constant bit rate}
\acro{CBS}{class based scheduling}
\acro{CC}{congestion control}
\acro{CCL}{common cell list}
\acro{CDF}{cumulative distribution function}
\acro{CDL}{clustered delay line}
\acro{CDMA}{code-division multiple access}
\acro{CIR}{channel impulse response}
\acro{CH}{channel hardening}
\acro{CHO}{conditional handover}
\acro{C-RAN}{cloud-based radio access network}
\acro{CL}{closed loop}
\acro{CLT}{central limit theorem}
\acro{CLPC}{closed loop power control} 
\acro{CN}{core network}       
\acro{CNR}{channel-to-noise ratio}
\acro{CP}{control plane}
\acro{CPA}{cellular protection algorithm}
\acro{CPICH}{common pilot channel}
\acro{CoMP}{coordinated multi-point}
\acro{CQI}{channel quality indicator}
\acro{CRE}{cell range expansion}
\acro{CRM}{constrained rate maximization}
\acro{CRN}{cognitive radio network}
\acro{C-RNTI}{cell radio network temporary identifier}
\acro{CRRM}{centralized/common radio resource management}
\acro{CRS}{cell-specific reference signal}
\acro{CS}{coordinated scheduling}
\acro{CSI}{channel state information}
\acro{CSI-RS}{channel state information reference signal}
\acro{CTS}{clear to send}
\acro{CU}{centralized unit}
\acro{CUE}{cellular user equipment}
\acro{CWND}{congestion window size}
\acro{D2D}{device-to-device}
\acro{D2N}{drone-to-network}
\acro{DAA}{drone antenna array}
\acro{DAG}{directed acyclic graph}
\acro{DBS}{drone mounted base station}
\acro{DC}{dual connectivity}
\acro{DCA}{dynamic channel allocation}
\acro{DE}{differential evolution}
\acro{DF}{decode and forward}
\acro{DFT}{discrete fourier transform}
\acro{DIST}{distance}
\acro{DL}{downlink}
\acro{DMA}{double moving average}
\acro{DMRS}{demodulation reference signal}
\acro{D2DM}{\acs*{D2D} mode}
\acro{DMS}{\acs*{D2D} mode selection}
\acro{DP}{data plane}
\acro{DPC}{dirty paper coding}
\acro{DRA}{dynamic resource assignment}
\acro{DSA}{dynamic spectrum access}
\acro{DSM}{delay-based satisfaction maximization}
\acro{DU}{distributed unit}
\acro{E2E}{end-to-end}
\acro{ECC}{electronic communications committee}
\acro{EDF}{earliest deadline first}
\acro{EE}{energy efficiency}
\acro{EFLC}{error feedback based load control}
\acro{EI}{efficiency indicator}
\acro{e-ICIC}{enhanced inter-cell interference coordination}
\acro{eMBB}{enhanced mobile broadband}
\acro{EM}{electromagnetic}
\acro{eNB}{evolved node B}
\acro{EXP}{exponential}
\acro{EPA}{equal power allocation}
\acro{EPC}{evolved packet core}
\acro{EPS}{evolved packet system}
\acro{E-UTRAN}{evolved universal terrestrial radio access network}
\acro{ES}{exhaustive search}
\acro{FCP}{fundamental counting principle}
\acro{FCA}{flow control algorithm}
\acro{FD}{full duplex}
\acro{FDD}{frequency division duplex}
\acro{FDM}{frequency division multiplexing}
\acro{FDMA}{frequency division multiple access}
\acro{FER}{frame erasure rate}
\acro{FIFO}{first in first out}
\acro{FoV}{field-of-view}
\acro{FF}{fast fading}
\acro{FR}{frequency range}
\acro{FRS}[FS]{fast-rat scheduling}
\acro{FS}{fast switching}
\acro{FSB}{fixed switched beamforming}
\acro{FST}{fixed \acs*{SNR} target}
\acro{FT}{fourier transform}
\acro{FTP}{file transfer protocol}
\acro{Fwd}{forwarding}
\acro{GA}{genetic algorithm}
\acro{GAP}{generalized assignment problem}
\acro{GAP-MQ}{generalized assignment problem with minimum quantities}
\acro{GATES}{generalized adaptive throughput-based efficiency-satisfaction trade-off}
\acro{GBR}{guaranteed bit rate}
\acro{GLR}{gain to leakage ratio}
\acro{gNB}{gNodeB}
\acro{GOS}{generated orthogonal sequence}
\acro{GP}{gaussian process}
\acro{GPL}{GNU general public license}
\acro{GPS}{global positioning system}
\acro{GRP}{grouping}
\acro{GSM}{global system for mobile communications}
\acro{GTEL}{wireless telecommunications research group}
\acro{HARQ}{hybrid automatic repeat request}
\acro{HBF}{hybrid beamforming}
\acro{HCPP}{hardcore point process}
\acro{HetNet}{heterogeneous network}
\acro{HD}{half duplex}
\acro{HF}{high-frequency}
\acro{HH}{hughes-hartogs}
\acro{HardH}[HH]{hard handover}
\acro{HMS}{harmonic mode selection}
\acro{HO}{handover}
\acro{HOL}{head of line}
\acro{HPBW}{half power beamwidth}
\acro{HSDPA}{high speed downlink packet access}
\acro{HSR}{high-speed railway}
\acro{HSPA}{high speed packet access}
\acro{HTTP}{hypertext transfer protocol}
\acro{IAB}{integrated access and backhaul}
\acro{ICMP}{internet control message protocol} 
\acro{ICI}{intercell interference}
\acro{ICIC}{inter-cell interference coordination}
\acro{ID}{identification}
\acro{i.i.d.}{independent and identically distributed}
\acro{IETF}{internet engineering task force}
\acro{IPC}{individual power constraint}
\acro{UID}{unique identification}
\acro{IID}{independent and identically distributed}
\acro{IIR}{infinite impulse response}
\acro{ILP}{integer linear problem}
\acro{IMT}{international mobile telecommunications}
\acro{INV}{inverted norm-based grouping} 
\acro{IoT}{internet of things}
\acro{IP}{internet protocol}
\acro{IPv6}{internet protocol version 6}
\acro{IRA}{integrated resource allocation}
\acro{IRS}{intelligent reflecting surface}
\acro{ISD}{inter-site distance}
\acro{ISI}{inter symbol interference}
\acro{ISM}{industrial, scientific and medical}
\acro{ITU}{international telecommunication union}
\acro{JOAS}{joint opportunistic assignment and scheduling}
\acro{JOS}{joint opportunistic scheduling}
\acro{JP}{joint processing}
\acro{JRAPAP}{joint rb assignment and power allocation problem}
\acro{JS}{jump-stay}
\acro{JSM}{joint satisfaction maximization}
\acro{KKT}{karush-kuhn-tucker}
\acro{KPI}{key performance indicator}
\acro{LAC}{link admission control}
\acro{LA}{link adaptation}
\acro{LBS}{location based service}
\acro{LC}{load control}
\acro{LOS}{line of sight}
\acro{LP}{linear programming}
\acro{LTE}{long term evolution}
\acro{LTE-A}{\acs*{LTE}-advanced}
\acro{LTE-Advanced}{\ac{LTE-A}}
\acro{LTE-R}{\ac{LTE} railway}
\acro{LSP}{large scale parameter}
\acro{MADRDPG}{multi-agent deep recurrent deterministic policy gradient}
\acro{MeNB}{master \acs*{ENB}}
\acro{M2M}{machine-to-machine}
\acro{MAC}{medium access control}
\acro{MANET}{mobile ad hoc network}
\acro{MEDS}{method of exact doppler spread}
\acro{MC}{modular clock}
\acro{MCP}{minimal cost power}
\acro{MCS}{modulation and coding scheme}
\acro{MDB}{measured delay based}
\acro{MDI}{minimum \acs*{D2D} interference}
\acro{MDSM}{modified delay-based satisfaction maximization}
\acro{MDU}{max-delay-utility}
\acro{METIS}{mobile and wireless communications enablers for the twenty-twenty information society \acs*{5G}}
\acro{MF}{matched filter}
\acro{MFAP}{mobile femtocell access point}
\acro{MG}{maximum gain}
\acro{MH}{multi-hop}
\acro{mIAB}{mobile \ac{IAB}}
\acro{MILP}{mixed integer linear programming}
\acro{MIMO}{multiple input multiple output}
\acro{MINLP}{mixed integer nonlinear programming}
\acro{MIP}{mixed integer programming}
\acro{MISO}{multiple input single output}
\acro{MIT}{mobility interruption time}
\acro{ML}{machine learning}
\acro{MLWDF}{modified largest weighted delay first}
\acro{MME}{mobility management entity}
\acro{MMF}{max-min fairness}
\acro{mmMAGIC}{millimetre-wave based mobile radio access network for fifth generation integrated communications}
\acro{MMRP}{maximizing the minimal residual power}
\acro{MMRP-LB}{maximizing the minimal residual power with lower bound}
\acro{MMSE}{minimum mean square error}
\acro{mMTC}{massive machine-type communications}
\acro{mmWave}{millimeter wave}
\acro{MN}{master node}
\acro{MOS}{mean opinion score}
\acro{MPF}{multicarrier proportional fair}
\acro{MPRP}{maximization of the product of the residual powers}
\acro{MRA}{maximum rate allocation}
\acro{MR}{maximum rate}
\acro{MRC}{maximum ratio combining}
\acro{MRN}{mobile relay node}
\acro{MRT}{maximum ratio transmission}
\acro{MRUS}{maximum rate with user satisfaction}
\acro{MS}{mode selection}
\acro{MSE}{mean squared error}
\acro{MCG}{master cell group} 
\acro{MSI}{multi-stream interference}
\acro{MT}{mobile termination}
\acro{MTC}{machine-type communication}
\acro{IMS}{\acs*{IP} multimedia subsystem}
\acro{MTSI}{multimedia telephony services over \acs*{IMS}}
\acro{MTSM}{modified throughput-based satisfaction maximization}
\acro{MU-MIMO}{multi-user multiple input multiple output}
\acro{MU}{multi-user}
\acro{Multi-CUT}{multi-cell and multi-user and multi-tier}
\acro{NAS}{non-access stratum}
\acro{NB}{node B}
\acro{nBS}{neighbor base station}
\acro{NCL}{neighbor cell list}
\acro{NCR}{network-controlled repeater}
\acro{NG}{next generation}
\acro{NGC}{next generation core network}
\acro{NLP}{nonlinear programming}
\acro{NLOS}{non-line of sight}
\acro{NMSE}{normalized mean square error}
\acro{NN}{neural network}
\acro{NOMA}{non-orthogonal multiple access}
\acro{NORM}{normalized projection-based grouping}
\acro{NP}{non-polynomial time}
\acro{NR}{new radio}
\acro{NRT}{non-real time}
\acro{NSA}{non-stand-alone}
\acro{NSPS}{national security and public safety services}
\acro{OBF}{opportunistic beamforming}
\acro{OFDMA}{orthogonal frequency division multiple access}
\acro{OFDM}{orthogonal frequency division multiplexing}
\acro{OFPC}{open loop with fractional path loss compensation}
\acro{O2I}{outdoor-to-indoor}
\acro{OL}{open loop}
\acro{OLPC}{open-loop power control}
\acro{OL-PC}{open-loop power control}
\acro{OM}[O\&M]{operational \& maintenance}
\acro{OMA}{orthogonal multiple access}
\acro{OPEX}{operational expenditure}
\acro{ORA}{orthogonal resource allocation}
\acro{ORB}{orthogonal random beamforming}
\acro{JO-PF}{joint opportunistic proportional fair}
\acro{OSI}{open systems interconnection}
\acro{PA}{power allocation}
\acro{PAIR}{\acs*{D2D} pair gain-based grouping}
\acro{PAPR}{peak-to-average power ratio}
\acro{P2P}{peer-to-peer}
\acro{PBCH}{physical broadcast channel}
\acro{PBS}{pico base station}        
\acro{PC}{power control}
\acro{PCI}{physical cell \acs*{ID}}
\acro{PDCP}{packet data convergence protocol}
\acro{PDF}{probability density function}
\acro{PDU}{protocol data unit}
\acro{PER}{packet error rate}
\acro{PF}{proportional fair}
\acro{PGRA}{probabilistic graph based resource allocation}
\acro{P-GW}{packet data network gateway}
\acro{PHY}{physical}
\acro{PL}{pathloss}
\acro{PLR}{packet loss ratio}
\acro{PRABE}{power and resource allocation based on quality of experience}
\acro{PRB}{physical resource block}
\acro{PROJ}{projection-based grouping}
\acro{PSD}{power spectral density}
\acro{PSDe}{positive semi-definite}
\acro{ProSe}{proximity services}
\acro{PS}{packet scheduling}
\acro{PSO}{particle swarm optimization}
\acro{PSS}{primary synchronization signal}
\acro{PTAS}{polynomial-time approximation scheme}
\acro{PTP}{point-to-point}
\acro{PZF}{projected zero-forcing}
\acro{QAM}{quadrature amplitude modulation}
\acro{QHMLWDF}{queue-HOL-MLWDF}
\acro{QoE}{quality of experience}
\acro{QoS}{quality of service}
\acro{QPSK}{quadri-phase shift keying}
\acro{QSM}{queue-based satisfaction maximization}
\acro{QuaDRiGa}{quasi deterministic radio channel generator}
\acro{RACH}{random access}
\acro{RAISES}{reallocation-based assignment for improved spectral efficiency and satisfaction}
\acro{RAN}{radio access network}
\acro{RA}{resource allocation}
\acro{RAP}{rb assignment problem}
\acro{RAT}{radio access technology}[long-plural-form={radio access technologies}]
\acro{RATE}{rate-based}
\acro{RAU}{remote antenna unit}
\acro{RB}{resource block}
\acro{RBG}{resource block group}
\acro{REF}{reference grouping}
\acro{RET}{remote electrical tilt}
\acro{RF}{radio frequency}
\acro{RL}{reinforcement learning}
\acro{RLC}{radio link control}
\acro{RLF}{radio link failure}
\acro{RM}{rate maximization}
\acro{RMa}{rural macro}
\acro{RMJ-SNR}{region of the minimum joint snr}
\acro{RMEC}{rate maximization under experience constraints}
\acro{RMSE}{root-mean-square error}
\acro{RNC}{radio network controller}
\acro{RND}{random grouping}
\acro{RNN}{recurrent neural network}
\acro{RRA}{radio resource allocation}
\acro{RRM}{radio resource management}
\acro{RSCP}{received signal code power}
\acro{RSRP}{reference signal received power}
\acro{RSRQ}{reference signal received quality}
\acro{RR}{round robin}
\acro{RIS}{reconfigurable intelligent surface}
\acro{RRC}{radio resource control}
\acro{RSSI}{received signal strength indicator}
\acro{RT}{real time}
\acro{RTS}{request to send}
\acro{RU}{resource unit}
\acro{RUNE}{rudimentary network emulator}
\acro{RV}{random variable}
\acro{Rx}{receiver}
\acro{RZF}{regularized zero-forcing}
\acro{SA}{subcarrier assignment}
\acro{SAC}{session admission control}
\acro{sBS}{serving base station}
\acro{SC}{small cell}
\acro{SCI}{side control information}
\acro{SCon}[SC]{single connectivity}
\acro{SCG}{secondary cell group}
\acro{SCM}{spatial channel model}
\acro{SCS}{subcarrier spacing}
\acro{SC-FDMA}{single carrier - frequency division multiple access}
\acro{SCRV}{spatially consistent random variable}
\acro{SD}{soft dropping}
\acro{SDAP}{service	data adaptation protocol}
\acro{S-D}{source-destination}
\acro{SDPC}{soft dropping power control}
\acro{SDM}{space-division multiplexing}
\acro{SDMA}{space-division multiple access}
\acro{SE}{squared error}
\acro{SF}{shadow fading}
\acro{SMDP}{semi-markov	decision problem}
\acro{SeNB}{secondary \acs*{ENB}}
\acro{SER}{symbol error rate}
\acro{SES}{simple exponential smoothing}
\acro{S-GW}{serving gateway}
\acro{SIC}{successive interference cancellation}
\acro{SelfIC}[SIC]{self interference cancellation}
\acro{SINR}{signal to interference-plus-noise ratio}
\acro{SLNR}{signal to leakage-plus-noise ratio}
\acro{SNN}{strictly non-negative}
\acro{SI}{satisfaction indicator}
\acro{SelfI}[SI]{self interference}
\acro{SIP}{session initiation protocol}
\acro{SISO}{single input single output}
\acro{SIMO}{single input multiple output}
\acro{SIR}{signal to interference ratio}
\acro{SM}{subcarrier matching}
\acro{SMA}{simple moving average}
\acro{SMSE}{spatial-mean-square error}
\acro{SN}{secondary node}
\acro{SNR}{signal to noise ratio}
\acro{SON}{self organizing networks}
\acro{SoA}{state-of-the-art}
\acro{SoS}{sum-of-sinusoids}
\acro{SORA}{satisfaction oriented resource allocation}
\acro{SORA-NRT}{satisfaction-oriented resource allocation for non-real time services}
\acro{SORA-RT}{satisfaction-oriented resource allocation for real time services}
\acro{SP}{signal processing}
\acro{SPF}{single-carrier proportional fair}
\acro{SR}{smart repeater}
\acro{ASR}{advanced \ac{SR}}
\acro{SRA}{sequential removal algorithm}
\acro{SRB1}{signaling radio bearer~1}
\acro{SRS}{sounding reference signal}
\acro{SSB}{synchronization signal block}
\acro{SSP}{small scale parameter}
\acro{SSS}{secondary synchronization signal}
\acro{ST}{spanning tree}
\acro{STTD}{space time transmit diversity}[long-plural-form={space time transmit diversities}]
\acro{SU-MIMO}{single-user multiple input multiple output}
\acro{SU}{single-user}
\acro{tBS}{target base station}
\acro{SVD}{singular value decomposition}
\acro{TCP}{transmission control protocol}
\acro{TDD}{time division duplex}
\acro{TDM}{time division multiplexing}
\acro{TDMA}{time division multiple access}
\acro{TDMed}{time division multiplexed}
\acro{TETRA}{terrestrial trunked radio}
\acro{TP}{transmit power}
\acro{TPC}{total power constraint}
\acro{TR}{technical report}
\acro{TTI}{transmission time interval}
\acro{TTR}{time-to-rendezvous}
\acro{TTT}{time-to-trigger}
\acro{TSM}{throughput-based satisfaction maximization}
\acro{TU}{typical urban}
\acro{TV}{television}
\acro{TVWS}{\acs*{TV} white space}
\acro{Tx}{transmitter}
\acro{UAV}{unmanned aerial vehicle}
\acro{UABSC}{user-assisted bearer split control}
\acro{UDP}{user datagram protocol}
\acro{UDN}{ultra-dense networks}
\acro{UE}{user equipment}
\acro{UBA}{\ac{UE}-\ac{BS} association}
\acro{UEPS}{urgency and efficiency-based packet scheduling}
\acro{UFC}{federal university of cear\'{a}}
\acro{ULA}{uniform linear array}
\acro{UL}{uplink}
\acro{UMa}{urban macro}
\acro{UMi}{urban micro}
\acro{UMTS}{universal mobile telecommunications system}
\acro{UP}{user plane}
\acro{UPN}{user provided network}
\acro{URA}{uniform rectangular array}
\acro{URI}{uniform resource identifier}
\acro{URLLC}{ultra-reliable low-latency communications}
\acro{URM}{unconstrained rate maximization}
\acro{VBR}{variable bit rate}
\acro{VET}{variable electrical tilt}
\acro{VMR}{vehicle-mounted relay}
\acro{VR}{virtual resource}
\acro{VoIP}{voice over \acs*{IP}}
\acro{WCDMA}{wideband code division multiple access}
\acro{WF}{water-filling}
\acro{WID}{wireless infrastructure drone}
\acro{Wi-Fi}{wireless fidelity}
\acro{WiMAX}{worldwide interoperability for microwave access}
\acro{WINNER}{wireless world initiative new radio}
\acro{WLAN}{wireless local area network}
\acro{WMPF}{weighted multicarrier proportional fair}
\acro{WP}{work package}
\acro{WPF}{weighted proportional fair}
\acro{WSN}{wireless sensor network}
\acro{WWW}{world wide web}
\acro{WFQ}{weighted fair queuing}
\acro{XIXO}{(single or multiple) input (single or multiple) output}
\acro{XPR}{cross-polarization power ratio}
\acro{ZF}{zero-forcing}
\acro{ZMCSCG}{zero mean circularly symmetric complex gaussian}

%% file: math.tex
\DeclareMathAlphabet{\mathppl}{T1}{ppl}{m}{it}
\DeclareMathAlphabet{\mathphv}{T1}{phv}{m}{it}
\DeclareMathAlphabet{\mathpzc}{T1}{pzc}{m}{it}

\newcommand{\ArgMax}[2]{\underset{#1}{\arg\max}\left\{#2\right\}}

\newcommand{\ArgMin}[2]{\underset{#1}{\arg\min}\left\{#2\right\}}
\newcommand{\Abs}[1]{\left\vert #1 \right\vert}

\newcommand{\Norm}[1]{\left\Vert #1 \right\Vert}

\newcommand{\Transp}[1]{{#1}^{\mathrm{T}}}

\newcommand{\Vt}[1]{\mathbf{\lowercase{#1}}}

\newcommand{\vtC}{\Vt{C}}

\newcommand{\vtR}{\Vt{R}}

\newcommand{\vtU}{\Vt{U}}
\newcommand{\vtV}{\Vt{V}}
\newcommand{\vtW}{\Vt{W}}